\documentclass{revtex4}
\usepackage{graphicx}
\usepackage{color}
\usepackage{amsfonts,amssymb,amsmath}
\usepackage{epsfig}
\usepackage{hyperref}

\renewcommand{\t}{\theta}
\renewcommand{\b}{\bar}
\newcommand{\ti}{\tilde}
\newcommand{\f}{\frac}

\newcommand{\nn}{\nonumber}

\newcommand{\La}{\Lambda}
\newcommand{\el}{\ell}
\newcommand{\h}{\hat}
\newcommand{\mrm}{\mathrm}
\newcommand{\be}{\begin{equation}}
\newcommand{\ee}{\end{equation}}
\newcommand{\ba}{\begin{eqnarray}}
\newcommand{\ea}{\end{eqnarray}}
\newcommand{\rh}{(r^2+a^2 \cos^2{\theta})}
\newcommand{\dr}{(r^2+a^2)(1-\frac{\Lambda}{3}r^2)- 2Mr }
\newcommand{\dt}{1+\frac{\Lambda}{3} a^2 \cos^2 \theta }
\newcommand{\s}{ 1 + \frac{\Lambda}{3} a^2 }

\begin{document}

\title{The Kerr-de Sitter Universe}

\author{Sarp Akcay}
\affiliation{ University of
Southampton}\email{sa18g09@soton.ac.uk} \affiliation{University of
Texas at Austin}
\author{Richard A. Matzner}

\affiliation{ Center for Relativity, University of Texas at Austin
\\ Texas Cosmology Center, University of Texas at Austin }

           \email{matzner2@physics.utexas.edu}


\begin{abstract}
It is now widely accepted that the universe as we understand it is
accelerating in expansion and fits the de Sitter model rather
well. As such, a realistic assumption of black holes must place
them on a de Sitter background and not Minkowski as is typically
done in General Relativity. The most astrophysically relevant
black hole is the uncharged, rotating Kerr solution, a member of
the more general Kerr-Newman metrics. A generalization of the
rotating Kerr black hole to a solution of the Einstein's equation
with a cosmological constant $\Lambda$ was discovered by Carter
\cite{DWDW}. It is typically referred to as the Kerr-de Sitter
spacetime. Here, we discuss the horizon structure of this
spacetime and its dependence on $\Lambda$. We recall that in a
$\La>0$ universe, the term `extremal black hole' refers to a black
hole with angular momentum $J > M^2 $. We obtain explicit
numerical results for the black hole's maximal spin value and get
a distribution of admissible Kerr holes in the ($\Lambda$, spin)
parameter space. We look at the conformal structure of the
extended spacetime and the embedding of the 3-geometry of the
spatial hypersurfaces. In analogy with Reissner-Nordstr\"{o}m -de
Sitter spacetime, in particular by considering the Kerr-de Sitter
causal structure as a distortion of the Reissner-Nordstr\"{o}m-de
Sitter one, we show that spatial sections of the extended
spacetime are 3-spheres containing 2-dimensional topologically
spherical sections of the horizons of Kerr holes at the poles.
Depending on how a $t=$ constant 3-space is defined these holes
may be seen as black or white holes (four possible combinations).

\keywords{Kerr, de Sitter, black holes}
\end{abstract}
\maketitle
\section{Introduction}\label{sec:intro}
The WMAP results of the last decade (\cite{WMAP4}, \cite{WMAP3},
\cite{WMAP2}, \cite{WMAP1}) have confirmed the current
cosmological paradigm: that our universe is given by the $\Lambda
CDM$ model with the dark energy component equivalent to a
cosmological ``constant" $\Lambda
> 0 $ making up almost three quarters of its total energy content
($ 0.728 \le \Omega_\Lambda \le 0.738 $ according to latest
analysis in \cite{WMAP4}). The first indication of a dark energy
component was obtained in 1998 when type Ia supernovae data were
analyzed and pointed toward an accelerating expansion of the
universe (\cite{SNI1}, \cite{SNI2}). While $\La$ made a negligible
contribution to the total energy density of the early universe,
the energy density of all other components (matter, cold dark
matter, radiation, neutrinos and gravitational waves) decreases as
the universe expands, so $\Lambda$ became (at some point in the
past) the dominant term in the total energy density. In the
distant future, the other components will be several orders of
magnitude smaller in density than $\La$, thus they will become
negligible. In this final state, the energy-momentum will be
completely given by:
\be T_{\mu\nu} = \La g_{\mu\nu}. \label{eq:Tmunu} \ee
%
The spatially homogeneous cosmological solution to Einstein's
equation ($G_{\mu\nu} = 8 \pi T_{\mu\nu} $) with this constant
$\La$ as the source term is the de Sitter (dS) spacetime. First
discovered by de Sitter in 1916 (\cite{dS1}, \cite{dS2},
\cite{dS3}), the de Sitter metric describes an empty universe
whose expansion is ever accelerating as time rolls forward. It is
best visualized as a 4-dimensional hyperboloid (see \cite{HE} for
nice illustrations)
\be -X_0^2+X_1^2+X_2^2+X_3^2+X_4^2 = r_C^2 \label{eq:4Dhyperbola}
\ee
embedded in 5-dimensional flat Minkowski spacetime
\be ds^2 = -dX_0^2+dX_1^2+dX_2^2+dX_3^2+dX_4^2 . \label{eq:5dMink}
\ee
To visualize the geometry of de Sitter spacetime in four
dimensions, we perform a coordinate transformation from
$(X_0,X_1,X_2,X_3,X_4) $ to $(T,\chi,\theta,\phi)$ coordinates
that satisfy Eq.(\ref{eq:4Dhyperbola}). The transformation
equations are
\ba r_C \sinh(T/r_C) = X_0 \ , & & \quad r_C \cosh(T/r_C) \cos\chi
= X_1, \nonumber \\ r_C \cosh(T/r_C) \sin \chi \cos\theta = X_2 \
, & & \quad r_C \cosh(T/r_C) \sin \chi \sin\theta \cos\phi = X_3,
\nn
\\ r_C \cosh(T/r_C) \sin \chi \sin\theta \sin\phi & =& X_4, \label{eqs:X_i2Glob}\ea
where $0\leq \chi, \theta \leq \pi$ and $ 0 \leq \phi \leq 2\pi$.
$r_C$ is a cosmological length scale that equals $ \sqrt{3
/\Lambda} $ (often referred to as the Hubble length). With these
substitutions, the de Sitter metric as given by
Eq.(\ref{eq:5dMink}) becomes \cite{HE}, \cite{Sch}
\be ds^2 = -dT^2 + r_C^2\cosh^2(T/r_C) \left[d\chi^2+ \sin^2
\chi \:(d\theta^2 + \sin^2 \theta \: d\phi^2) \right] \ .
\label{eq:dSmetricTchi}\ee
As can be seen from the metric of Eq.(\ref{eq:dSmetricTchi}), the
geometry of de Sitter spacetime is that of a 3-sphere which
contracts to a minimum size at $T=0$ and then expands
hyperbolically in time. The coordinates $(T,\chi,\theta,\phi)$ are
called the global coordinates of de Sitter spacetime because they
cover the entire 4-dimensional spacetime manifold. The
$T=constant$ surfaces are 3-spheres of constant positive
curvature.

There are other choices for coordinates in dS spacetime. One such
system of coordinates $(\hat{t}, \hat{x}, \h{y}, \h{z})$ is the
so-called {\it conformally flat} (\cite{HE}) slicing of the
4-dimensional hyperboloid given by the following relations:
\be \h{t} =r_C \ln\left(\f{X_0+X_1}{r_C}\right),\quad \h{x} =
\f{r_C X_2}{X_0+X_1}, \quad \h{y} = \f{r_C X_3}{X_0+X_1},
\quad\h{z} = \f{r_C X_4}{X_0+X_1}, \label{eq:Glob2flat} \ee
which give us the following for the de Sitter metric
\be ds^2 = -d\h{t}^2 + \mrm{e}^{2\h{t}/r_C}
\left(d\h{x}^2+d\h{y}^2+d\h{z}^2\right)\ .\label{eq:dSmetric_flat}
\ee
Note that $\h{t} = constant$ hypersurfaces are flat in this
coordinate system. However, because $\h{t}$ is ill-defined for
$X_0+X_1 \leq 0 $, the conformal coordinates cover only half of
the hyperboloid.

\begin{figure*}
\includegraphics[height=4.7cm]{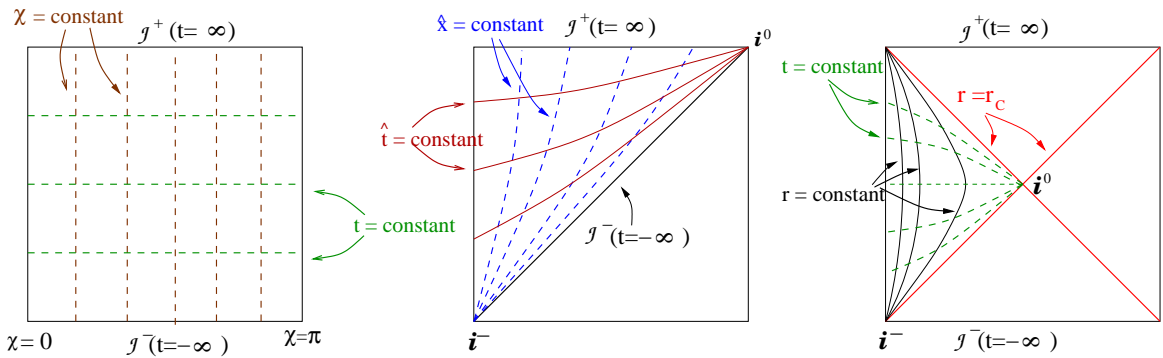}
\caption{de Sitter spacetime presented in three different
coordinate systems. From left to right, we have the Penrose
diagrams with global, hyperbolic and static coordinates. As can be
seen in the figure, global coordinates cover the entire manifold
whereas static coordinates only cover a quarter of it.
$\mathcal{J}^\pm $ represent future($+$) and past ($-$) null
infinities. A peculiar property of de Sitter spacetime is that
null future infinities $ \mathcal{J}^\pm$ can be spacelike
surfaces \cite{HE} where all timelike and null lines end up. $
{\mathbf{\it i}}^- $, $ \mathbf{\it {i}}^0 $ represent past
timelike ($0 \le r \le r_C $) and spacelike infinities,
respectively (see \cite{HE}, chapter 5).}\label{fig:dS}
\end{figure*}

Finally, we can choose to use another set of coordinates $
(t,r,\theta,\phi)$ given by the following \cite{HE}, \cite{Sch}:
%
\ba X_0 = \sqrt{r_C^2-r^2} \sinh{t/r_C}, \quad X_1 =
\sqrt{r_C^2-r^2} \cosh{t/r_C}, \nn \\ X_2 = r \sin\theta \cos\phi,
\quad X_3= r \sin\theta \sin\phi, \quad X_4 = r \cos\theta \ .
\label{eqs:Glob2static} \ea
This set of coordinates is known as {\it static coordinates}
because of the particular form that the dS metric acquires:
\be ds^2 = -\left(1-\f{\La}{3}r^2\right) dt^2 +
\f{dr^2}{1-\f{\La}{3}r^2} + r^2 \left(d\theta^2 + \sin^2\theta
d\phi^2 \right). \label{eq:dSmetric_static} \ee
Clearly, $t=constant $ slices in this coordinate system are
invariant under time translations and reversals. In other words,
the 3-geometry of the hypersurfaces is constant in time i.e. {\it
static}. As can be seen from the right hand panel of Fig.
\ref{fig:dS}, there is clearly a null boundary in the static
slicing of de Sitter spacetime. Physically, this null surface
corresponds to the cosmological horizon of de Sitter spacetime
(where the expansion rate of spacetime exceeds the speed of
light). This horizon is located at $r=r_C$ where $r_C$ is the root
of $g_{rr}^{-1} = 0 $ given in Eq.(\ref{eq:dSmetric_static}): $r_C
= \sqrt{3/\Lambda} $. This is the same $r_C$ as the one in
Eqs.(\ref{eq:4Dhyperbola}) --- (\ref{eqs:Glob2static}). Although
the significance of $r_C$ as a horizon only becomes apparent in
static coordinates, it clearly is the cosmological characteristic
length scale for de Sitter spacetime in all coordinate systems.
Static coordinates are especially relevant since the Kerr-de
Sitter metric is usually presented in terms of them, and the
horizon structure of the Kerr-de Sitter spacetime is most
transparent in these coordinates.

There has been a sporadic influx of interest on the Kerr-de Sitter
solution since the 1970s. Most of this has been in the context of
singularity theorems, string theory (mostly focused on
Kerr-Anti-de Sitter) and trying to come up with physical
interpretations for solutions to Einstein equation with three or
more parameters. In this article, we provide a clear and concise
introduction to this special solution. We also present numerical
results that quantify the properties of Kerr black holes in this
spacetime. We show that in a $\La > 0 $ universe, the black hole
spin can exceed the $ a = M $ bound and that for $ \La > 1/9
M^{-2}$ non-rotating black holes cannot exist. Finally, we clarify
and provide some foundations for understanding what may be a well
known but not so well grasped peculiarity of
Reissner-Nordstr\"{o}m-de Sitter and Kerr-de Sitter solutions: The
time reversal of the evolution neighboring regions that are
separated by the null horizon $r_C$. Although this time reversal
is often mentioned in the literature, how it comes about has
hardly ever been explained.

The remainder of the paper is organized as follows. In Section
\ref{sec:SdS} we introduce Schwarzschild-de Sitter spacetime. This
is our `warm-up' cosmological model for Kerr-de Sitter. In section
\ref{sec:KdS}, we present a thorough study of the Kerr-de Sitter
universe and present its peculiar properties. One of these is the
possibility that the black hole spin parameter $a$ may exceed the
black hole mass parameter $M$. An extreme Kerr black hole in
Kerr-de Sitter will have $a>M$; we provide numerical examples in
section \ref{sec:KdS}. In section \ref{sec:Globe}, we will
investigate the conformal structure of the Kerr-de Sitter universe
and the extended spacetime. We extend the correspondence between
Reissner-Nordstr\"{o}m-de Sitter (RNdS) and Kastor-Trashen (KT)
spacetimes to an approximate correspondence between Kerr-de Sitter
and KT spacetimes. With this we show that the extended Kerr-de
Sitter spacetime can be physically interpreted as a 3-sphere
containing a Kerr hole and a counter-rotating Kerr hole located
antipodally from one another. Thus the Kerr-de Sitter spacetime
can be extended to two holes analogously to the extension of the
charged RNdS case (first interpreted to represent two oppositely
charged holes in \cite{MM}).
\section{Schwarzschild-de Sitter Universe}\label{sec:SdS}
Schwarzschild-de Sitter (SdS) spacetime is a spherically symmetric
uncharged black hole solution to Einstein's equation with the
cosmological constant $\La$ as the source term
(Eq.(\ref{eq:Tmunu})) \cite{Ko}, \cite{Weyl}, \cite{Tr},
\cite{Sch}. It is described by two parameters: the black hole mass
$ M $ and the cosmological constant $\La$. Both of these
parameters are taken to be positive in order to obtain the SdS
solution. In static coordinates, the SdS metric is given by the
following expression \cite{To}:
\be ds^2 = -\left(1-\f{2M}{r}-\f{\La r^2}{3}\right) dt^2 +
\f{dr^2}{\left(1-\f{2M}{r}-\f{\La r^2}{3}\right)} + r^2
d\Omega_2^2,   \label{eq:SdS_metric} \ee
where $ d\Omega_2^2$ is the metric on a 2-sphere. To determine the
position of the event horizon(s) in this spacetime, we must find
where the expansion, $\Theta_{(\ell)}$, of the outgoing null
geodesics $\ell^\mu$ equals zero. Technically speaking, this
method gives the locations of the apparent horizons, but since the
spacetime is static, the apparent and the event horizons coincide.
A quick computation (see Appendix A2 in \cite{Ash} for details)
shows that the expansion is proportional to $g_{rr}^{-1}$ of
Eq.(\ref{eq:SdS_metric}). So, the roots of
\be g_{rr}^{-1} = 1-\f{2M}{r}-\f{\La r^2}{3} = 0
\label{eq:cubicSdS} \ee
give us the locations of the horizons in this spacetime. This
equation is a truncated cubic which generally yields three complex
roots. The condition for all three roots to be real is $ 9M^2 \La
< 1 $ ($\La =(9M^2)^{-1}$ corresponds to the degenerate horizon
case where two real roots coincide). Even with this condition,
only two of these three roots are positive. Physically, the
smaller positive root corresponds to the Schwarzschild black hole
horizon $r_H$ and the larger root to the cosmological horizon
$r_C$. The explicit expressions for the roots can be presented in
the following form \cite{GN} (also see \cite{HNS} and its Fig.1):
\ba r_1 & = & \f{2}{\sqrt{\La}} \cos\f{\theta}{3}, \nn \\
r_2 & = & \f{2}{\sqrt{\La}} \cos
\left[\f{\theta}{3}+\f{2\pi}{3}\right], \nn \\
r_3 & = & \f{2}{\sqrt{\La}} \cos
\left[\f{\theta}{3}+\f{4\pi}{3}\right], \nn \ea
where $\theta$ is defined by $ \cos\theta = -3\sqrt{\La}M $. With
the reality condition $9M^2 \La <1$, $0 \ge \cos\theta > -1 $ for
$M, \La \ge 0 $. Thus, we have $ \pi/2 < \theta< \pi $. In this
range for $\theta$, the positive roots are $r_1$ and $r_3$ with
$r_3 < r_1$; $r_2$ is negative. So, $r_3$ is the black hole
horizon $r_H$ and $r_1$ is the de Sitter horizon $r_C$. The
reality condition has the physical interpretation of imposing an
upper mass limit $M_{max} = \La^{-1/2}/3$ on the Schwarzschild
black hole; in other words, requiring that the black hole radius
$r_H$ be roughly smaller than the size of the de Sitter horizon $
r_C$. Obviously, at $ 9M^2 \La = 1 $ i.e. $M=M_{max} $, the two
horizons coincide ($ r_H = r_C = \La^{-1/2} $).

\begin{figure*}
\includegraphics[height=7cm]{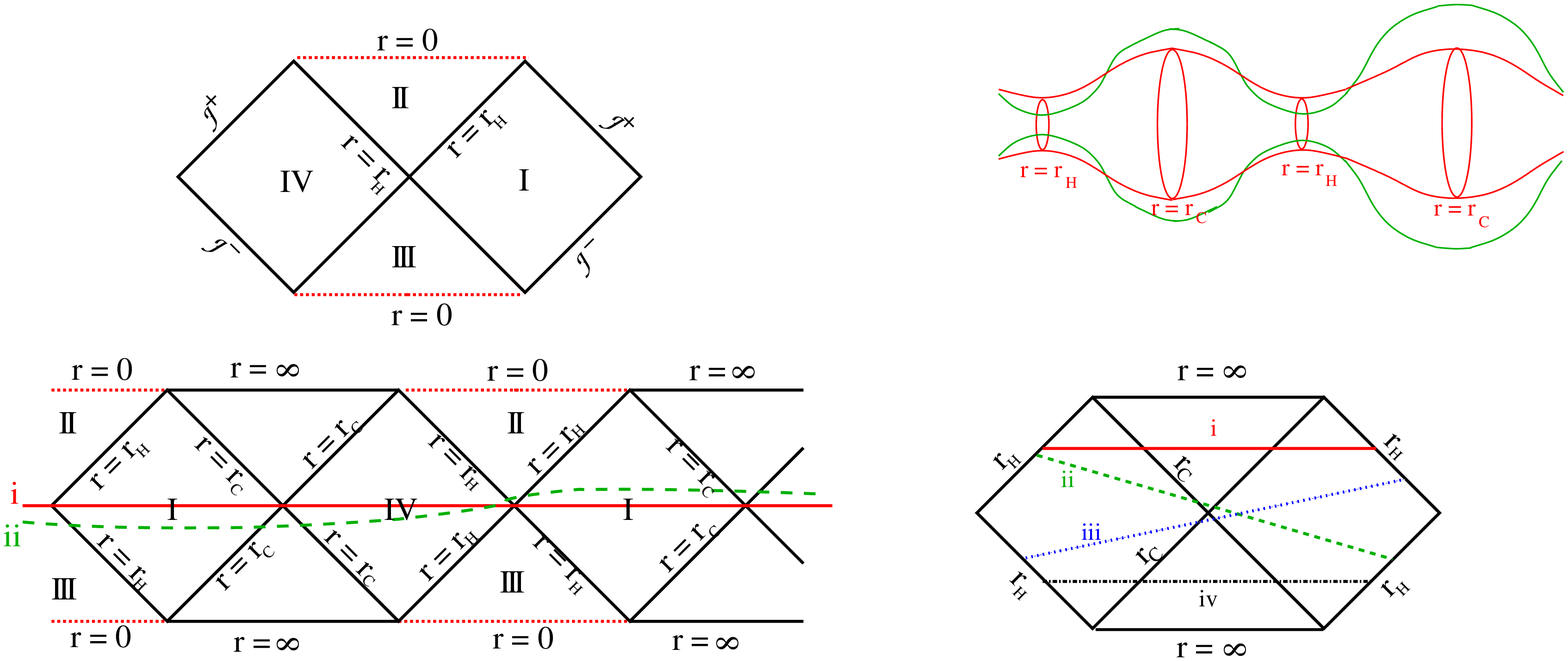}
\caption{The Penrose diagrams for Schwarzschild (upper-left) and
Schwarzschild-de Sitter (lower-left) spacetimes. The bottom
diagram continues ad infinitum to the right and to the left with
infinitely many curvature singularities ($r=0$) and spacelike
infinities ($r=\infty$). $\mathcal{J}^\pm$ label the future and
past null infinities . Region I in SdS spacetime is analogous to
region I of Schwarzschild spacetime where the Killing vector
$\partial/
\partial t $ is timelike and future directed. It is timelike, but
past directed in region IV and spacelike in regions II and III.
The horizontal solid/red (labelled i) curve is a time-symmetric
cut of the spacetime, whereas the dashed/green (ii) curve
represents a more generic spacelike hypersurface. The ``beads on a
string'' on the upper-right of the figure show the geometry along
hypersurfaces i and ii. Of course, there is no string, but simply
black/white hole throats connecting (not causally) each bead
universe to the next. Finally in the lower-right panel, we slice a
2-hole, 1-bead section of the SdS spacetime to illustrate the
different scenarios involving white/black holes in the 3-geometry
of the hypersurfaces. We pick 4 different hypersurfaces: red,
solid (i); green, dashed (ii); blue, dotted (iii); black,
dash-triple-dotted (iv). These slicings respectively give $ BH-BH,
BH-WH, WH-BH, WH-WH $ within the chosen hypersurface.}
\label{fig:SdS}
\end{figure*}

The details of how one obtains the conformal structure of SdS
spacetime can be found in \cite{GN}. The Penrose-Carter diagram
for the SdS spacetime \cite{GH} is displayed in the lower-left
panel of Fig. \ref{fig:SdS}. Unlike in Schwarzschild spacetime,
two separate coordinate charts are needed to cover the SdS
spacetime. One proceeds by introducing Kruskal type null
coordinates in the vicinity of the black hole horizon and
Hawking-Gibbons coordinates \cite{GH} in the vicinity of the
cosmological horizon. Furthermore, there are now infinitely many
regions where the Killing vector $\partial/
\partial t $ is timelike as opposed to only two regions in regular Schwarzschild
spacetime. There is an infinite sequence of singularities $ r= 0$
and spacelike infinities $r=\infty$ in the conformal
representation of SdS universe. So, on any spacelike hypersurface
that we may pick in this extended spacetime, we would get a
picture similar to that of an infinite number of beads on a wire.
However, in this case, the beads are actually connected (but not
causally) by throats of different sizes, which depend on how far
into the black hole region ($ r < r_H $) a hypersurface crosses.
The same goes for the maximum radii of the beads as a given
hypersurface can go arbitrarily far beyond $ r=r_C$ before
`exiting' through the companion cosmological horizon. We
illustrate the geometries of two different hypersurfaces in the
upper-right portion of Fig. \ref{fig:SdS}.

Furthermore, the `color' of the `black' hole that one observes is
determined by where the observer's spacelike slice intersects a
given horizon $r_H $. Depending on the location of the
intersection points, one may end up with any one of the following
four combinations for the two trapped regions: $ Black \ Hole\:
(BH)-BH,\: White \ Hole \:(WH)-WH, \:BH-WH,\: WH-BH $ (We define a
BH to be a the region inside a future event horizon; a WH as the
region inside a past event horizon). The various pairings between
BHs and WHs are shown in the lower-right panel of Fig.
\ref{fig:SdS} where we slice the $ r_H < r < r_C $ region of
spacetime with four different hypersurfaces extending from one
hole to the next.

\section{Kerr-de Sitter Spacetime}\label{sec:KdS}
The solution to Einstein's equation with cosmological constant
describing a black hole with spin $a$ is called the Kerr-de Sitter
(KdS) solution. It was found by Carter and was first published in
the 1973 {\it Les Houches} lectures edited by DeWitt \& DeWitt
\cite{DWDW}. Later, it was discovered that this solution is a
special case of the general Pleba\'{n}ski-Demia\'{n}ski family of
metrics \cite{PD}, \cite{Exact}, \cite{GF1}, \cite{GF2}. The
Pleba\'{n}ski-Demia\'{n}ski solution is the most general solution
for a Petrov Type D spacetime.
Kerr-de Sitter happens to be a very restricted case of the general
Plebansk\'{y}-Demiansk\'{y} solution with zero NUT charge,
acceleration, electric and magnetic charges. In the
Boyer-Lindquist like coordinates used by Carter, the Kerr-de
Sitter metric is as follows
\ba ds^2 & = & \rh \left[\f{dr^2}{\Delta_r} + \f{d\theta^2}{\dt}
\right] + \sin^2\theta\:\f{\dt}{\rh} \left[\f{a dt - (r^2+a^2) d
\phi}{\s}\right]^2 \nonumber \\ & \quad & -
\f{\Delta_r}{\rh}\left[\f{dt - a \sin^2\t \:d\phi}{\s}\right]^2,
\label{eq:KdS_metric1} \ea
where \be \Delta_r = \dr = r^2-2Mr+a^2-\f{\La r^2}{3}(r^2+a^2).
\label{eq:Delta_r} \ee
The constant $\La$ has the opposite sign here compared to Carter's
original version in \cite{DWDW}; Hawking \& Gibbons \cite{GH}
agree with our choice of sign for de Sitter. This choice of sign
is consistent with our sign of $\Lambda$ for SdS metric (see
Eq.(\ref{eq:SdS_metric})). As is well known, the $ a\rightarrow 0,
M \ne 0 $ limit of KdS is SdS.

We will write Eq.(\ref{eq:KdS_metric1}) in a more compact form:
\be ds^2 = \rho^2 \left[\f{dr^2}{\Delta_r} +
\f{d\t^2}{\Delta_{\t}}\right] + \f{\Delta_\t \sin^2\t}{\rho^2}
\left[a \f{dt}{\Xi} - (r^2+a^2) \f{d\phi}{\Xi}\right]^2 -
\f{\Delta_r}{\rho^2}\left[\f{dt}{\Xi}- a \sin^2\t
\:\f{d\phi}{\Xi}\right]^2 \label{eq:KdS_metric2}, \ee
where
\ba  \Delta_\t  & = & \dt, \nonumber \\
    \rho^2 & = & \rh, \nonumber \\
    \Xi & = & \s . \ea
Carter included the factor of $\Xi $ to ensure that there are no
conical singularities on the spatial axis $\theta=\{0,\pi\}$. We
show the details of how this works in appendix \ref{sec:app1}. If
desired, one can rescale the Boyer-Lindquist time coordinate $t$
by a factor of $\Xi^{-1}$; the version of the KdS metric presented
in \cite{GLPP} indeed embodies this slight modification.

To see that this is indeed asymptotically de Sitter, consider the
$M=0$ limit of Eq.(\ref{eq:KdS_metric1}):
\ba ds^2 & = & \rh \left[\f{dr^2}{(r^2+a^2)(1-\Lambda r^2/3)} +
\f{d\theta^2}{\dt} \right] \nn \\ &\quad &+
\sin^2\theta\:\f{\dt}{\rh} \left[\f{a dt - (r^2+a^2) d
\phi}{\s}\right]^2 \nonumber \\ & \quad & - \f{(r^2+a^2)(1-\Lambda
r^2/3)}{\rh}\left[\f{dt - a \sin^2\t \:d\phi}{\s}\right]^2
\label{eq:KdS_M0limit}. \ea
It is not so obvious from Eq.(\ref{eq:KdS_M0limit}) that what one
has is regular de Sitter spacetime. This is because Carter's
Boyer-Lindquist type coordinates need to be ``untwisted'' in order
for us to really see that Eq.(\ref{eq:KdS_M0limit}) gives the dS
metric. The untwisting is realized by the following coordinate
transformation:
\ba T = t/\Xi, & & \qquad \b{\phi} = \phi - a \Lambda t/3\Xi,
\nonumber
\\ y \cos\Theta & = & r \cos\t, \nonumber \\
\quad y^2 & = & \f{1}{\Xi}\left[r^2 \Delta_\t + a^2 \sin^2 \t
\right] \label{eq:Hawk_coord_tran} . \ea
By doing the algebra, one can show that these coordinate
transformations yield the usual form of the de Sitter metric in
($T, y, \Theta, \b{\phi} $) coordinates:
\be ds_{dS}^2 = - (1- \Lambda y^2/3) dT^2 + \f{1}{1-\Lambda y^2/3}
dy^2+ y^2 (d\Theta^2 + \sin^2 \Theta \: d\b{\phi}^2).
\label{eq:dSmetric_inY} \ee
It is also interesting to note that one can cast the KdS spacetime
into a Kerr-Schild like form. In general, the Kerr-Newman vacuum
solution to Einstein's equation can be written in a special form
called the Kerr-Schild form of the metric. This form is given by
(\cite{Ch},\cite{MTW}, \cite{Po}, \cite{Ho}),
\be g_{\mu \nu} = \eta_{\mu \nu} + 2 H k_{\mu} k_{\nu},
\label{eq:KSmetric} \ee
where $ H $ is a function of spacetime coordinates and for the Kerr solution
is proportional to the constant $M$: $H = 2Mr/\rho^2$. $ \eta_{\mu
\nu} $ is the Minkowski metric of flat spacetime and $ k^{\mu} $
is a null vector with respect to both $ g_{\mu \nu} $ and $
\eta_{\mu \nu} $. Of course, here we are not dealing with the
Minkowski metric; the $M \rightarrow 0 $ limit of Kerr-de Sitter
metric is simply the de Sitter metric as was shown above. But in analogy
to Eq.(\ref{eq:KSmetric}), we write
\be ds^2 = ds_0^2 + H (k_\mu dx^\mu)^2 \label{eq:KdS_KSmetric},
\ee
where $ H = 2Mr/ \rho^2 $ and $ds_0^2$ is the background de-Sitter
metric. We can express this background in Kerr-Schild (KS)
coordinates $(\tau, r, \theta, \Phi)$:
\be ds_0^2 = -\f{\Delta_\theta}{\Xi} (1-\La r^2/3) d\tau^2 +
\f{\rho^2}{(r^2+a^2)(1-\La r^2/3)} dr^2 +
\f{\rho^2}{\Delta_\theta} d\theta^2 + \f{r^2+a^2}{\Xi}\sin^2\theta
d\Phi^2 . \label{eq:KdS_KSflat} \ee
The null vector equals
\be k_\mu dx^\mu = \f{\Delta_\theta}{\Xi} d\tau +
\f{\rho^2}{(r^2+a^2)(1-\La r^2/3)} dr - \f{a\sin^2\theta}{\Xi}
d\Phi . \label{eq:KdS_KS_k} \ee
The coordinate transformation that connects KS and BL coordinates
is given in \cite{GLPP} and is as follows:
\be d\tau = dt + \f{2Mr}{(1-\La r^2/3)\: \Delta_r} dr, \quad d\Phi
= d\phi -\f{\La}{3} a  dt + \f{2Mr}{(r^2+a^2)\: \Delta_r} dr .
\label{eq:KStoBL} \ee
Applying this coordinate transformation to obtain Carter's form of
the KdS metric in BL coordinates (Eq.(\ref{eq:KdS_metric2})) is
somewhat tedious, but straightforward. At first, the resulting
metric in BL coordinates $(t,r,\theta,\phi)$ appears to be
different:
\ba ds^2 & = & -\f{\Delta_\theta}{\Xi} (1-\La r^2/3) dt^2 +
\f{2Mr}{\rho^2}\left[dt-\f{a\sin^2\theta}{\Xi} d\phi \right]^2 \nn \\
& & + \f{r^2+a^2}{\Xi} \sin^2\theta \left(d\phi - \f{\La a}{3}
dt\right)^2
\f{\rho^2}{\Delta_r} dr^2 + \f{\rho^2}{\Delta_\theta}
d\theta^2. \label{eq:KdS_KS2BLmetric} \ea
However, with a little work, one can show that this is indeed
Carter's metric (Eq.(\ref{eq:KdS_metric1}) with the one
difference in the scaling of the time coordinate by a factor of $\Xi = 1 +
\La a^2/3 $.

We are interested in understanding the horizon structure of this
spacetime. We look for the locations of the apparent horizons in
the same way that we have done in section \ref{sec:SdS}. First, we
find the principal null vectors ($\ell^\mu, n^\mu $) of the
spacetime. The Pleba\'{n}ski-Demia\'{n}ski metrics are of Petrov
Type D, which means that there are only two principal null vectors
\cite{Ch}, \cite{Exact}. The $\ell^\mu$ and $n^\mu $ are defined
only up to an overall functional factor; we impose the standard
relative normalization $ \ell_\mu n^\mu = -1$, which still allows
for a scaling of $\el^\mu$ with an inverse scaling of $n^\mu$. We
then compute the expansion of the outgoing null rays $\ell^\mu$ on
the 2-surface that is transverse to the 1+1 dimensional spacetime
spanned by $\ell^\mu$ and $n^\mu$; where this expansion equals
zero is the location of the apparent horizon. Mathematically
speaking, we solve $\Theta_{(\ell)} \equiv q^\mu_{\ \nu}
\nabla_\mu \ell^\nu = 0$ with the metric of the transverse
2-surfaces given by $q_{\mu\nu} = g_{\mu\nu} + (\ell_\mu
n_\nu+n_\mu \ell_\nu)/(-\ell_\alpha n^\alpha)$. For KdS spacetime,
in Carter's BL type coordinates, the outgoing null vector
$\ell^\mu$ is
\be \ell^\mu  =  \sqrt{\f{\Delta_r}{2 \rho^2}}
\left(\f{r^2+a^2}{\Delta_r} \Xi, 1, 0, \f{a}{\Delta_r} \Xi
\right). \label{eq:lmu_inBL}\ee
%
%
Notice that $\el^\mu$ is a Killing vector of the spacetime. For
the expansion, we obtain
%
\be \Theta_{(\el)} \propto 3r \; \Delta_r . \label{eq:expansion3}
\ee
Here, we deliberately left out coefficients (all positive) that tell us nothing
about the location of apparent horizons (where $\Theta_{(\el)}=0 $).
%
\ba \Theta_{(\el)} = 0 \quad \mrm{where} \quad && \ \Delta_r  \: = \: 0 ; \nonumber \\
& & \dr = 0; \nonumber \\
& & (r-r_+) (r-r_-) (r-r_C) (r-r_{--}) = 0 \ . \label{eq:Delta_roots}
\ea
Here, $ r_{\pm} $ are the Kerr black hole horizons, and $ r_C $ is
the cosmological horizon. $r_{--}<0$ is another cosmological
horizon ``inside" the singularity at $r=0$.
Eq.(\ref{eq:Delta_roots}) yields four real roots only for certain
values of $ \Lambda = \Lambda(M,a) $ or $M= M(a,\Lambda)$. To
determine what the limiting values are, we impose the restriction
on the quartic formula \cite{HMF} that all four roots be real.
This translates to an upper bound for $ M $ that we call $ M_{max}
$ which is determined by the following equation:
%
%
%
\be \left[\f{1}{\Lambda}\left(1-\f{\Lambda}{3}a^2\right) - 4 a^2
\right]^3 \quad = \quad \f{1}{\Lambda}
\left[\left(1-\f{\Lambda}{3}a^2\right)\left[\f{1}{\Lambda}
\left(1-\f{\Lambda}{3}a^2\right)^2 + 12 a^2\right]-18M_{max}^2
\right]^2. \label{eq:M_max} \ee
Only for $ M < M_{max} $ do we get three distinct positive roots
for $ \Delta_r $, which give us the two black hole horizons $
r_{\pm} $ and the cosmological horizon $ r_C $.
Eq.(\ref{eq:M_max}) is rather complicated to solve analytically
without assigning specific values for $ M $ and $ a $. However, we
can at least check this condition at the $ a = 0 $ limit, which
gives
\be M < M_{max} = \f{1}{3 \sqrt{\La}}, \label{eq:M_max_Sch} \ee
agreeing with the Schwarzschild-de Sitter result. From a physical
perspective, one sees that an upper bound on the black hole mass
$M$ simply reflects the fact that since the cosmology induces a
relative recession speed between two radially separated points that
increases with distance, the horizon of a larger black
hole would be torn apart by the cosmic expansion.

We prefer not to display the solutions for the roots of $ \Delta_r
$ explicitly, since writing down the solutions of a quartic
equation using the quartic formula is not always enlightening. By
assigning values to the black hole's mass $M$ and spin $a$, we can
make the solutions to the quartic equation look less crowded
without loss of generality. For simplicity, we set $ M $ equal to
1 and scale the values for $a$ and $\Lambda$ accordingly (recall
that $\La$ has dimensions of (length)$^{-2}$). A very interesting
case is determining the horizon locations for Kerr when $ a =M $.
One would naively expect that the black hole roots $ r_- $ and $
r_+ $ meet at a single location as this is the case in Kerr
spacetime. But plotting this quartic function in the proper range
for $\La$ shows that the function still possesses three separate,
positive roots. Solving Eq.(\ref{eq:M_max}) for $\La_{max}$ with
$a=M$ yields
\be \ti{\La}_{max} \approx 0.1528377 \ , \label{eq:L_max_for_a_1}
\ee
where we have introduced the dimensionless cosmological constant
$\ti{\La} \equiv  M^2 \La $. Similary, for the spin, $ \ti{a}
\equiv  a/M $ and radial distance $\ti{r} \equiv r/M $. At $\La =
\La_{max} $, the outer roots $r_+$ and $r_C$ merge, which
physically corresponds to what is called \emph{the extreme (or
marginal) naked singularity case} in \cite{BH}. The double root is
a degenerate cosmological horizon and the smaller root $r_-$ is
the usual Cauchy horizon. Any value we pick for $\ti{\La}$ less
than $\ti{\La}_{max} \approx 0.1528377 $ should give us an
asymptotically de Sitter spacetime with a regular Kerr black hole
in it. Let us pick $\ti{\La}=0.1$ for example (This should read
$\La M^2 = 0.1$ if we restore the dimensions). Solving the quartic
for $ \Delta_r $ explicitly with $ M = a  $ and $\ti{\La}=0.1$
yields \emph{three} positive roots (not two) to
\be \Delta_r = r^2 - 2r + 1 - \f{0.1}{3} r^2 (r^2 + 1) = 0,
\label{eq:Delta_r_extremal} \ee
which are
\be \ti{r}_- \approx 0.8097645665 , \quad \ti{r}_+ \approx
1.485880984 ,\quad \ti{r}_C \approx 3.975723107 .
\label{eq:roots_a_1} \ee
Clearly, the $r_\pm$ are far from merging for this value of $\La$,
which tells us that the $a=M$ value no longer represents the
extremal Kerr black hole. Since we know that the $\La\rightarrow
0$ limit of KdS is regular Kerr spacetime, if we pick a much
smaller value for $\La$, we should get $r_-\approx r_+$. For
example, for $\ti{\La} = 10^{-4}$, one gets
\[ \ti{r}_- \approx 0.9919334965, \quad \ti{r}_+\approx 1.008266559 . \]
These numerical results are simply counter-examples showing that
$a=M$ case does not yield an extremal Kerr black hole in Kerr-de
Sitter spacetime. Then, what value of $a=a_{max}$ makes the black
hole extremal? For extremality, the black hole horizons should
coincide i.e. $r_+ = r_-$. This implies that the surface gravity
$\kappa$ of the Kerr-de Sitter black hole equals zero. Since
$r_\pm$ are roots of $\Delta_r$, a fourth-order polynomial in $r$,
the only way one can have $r_+=r_-$ is if the two roots merge at a
local extremum of $\Delta_r$. Thus, the following condition needs
to be satisfied at $a=a_{max}$:
\be \partial_r \Delta_r(r=r_\pm) = 0 \label{eq:amax_cond}. \ee
Since $a_{max}$ depends on $\La$, we proceed by first specifying a
value for $\La$ (setting up the cosmology) then solving the
$\Delta_r = 0$ quartic equation (Eq.(\ref{eq:Delta_roots})) for
the middle two roots $r_-$ and $r_+$. At this point, since the
spin $a$ is at a given arbitrary initial value, we do not expect
that $r_+=r_-$. To ensure this, we evaluate $ \partial_r \Delta_r
|_{r=r_\pm} $ using $a$ as a free parameter. Where this expression
equals zero we conclude that the two roots $r_-$ and $r_+$ have
merged to form an extremum of $\Delta_r$; and as stated above,
this only happens at $a=a_{max}$. We use standard Newton-Raphson
iteration where we input the values for $M$ and $\La$ a priori
then start with
\begin{figure*}
\includegraphics[height=10cm]{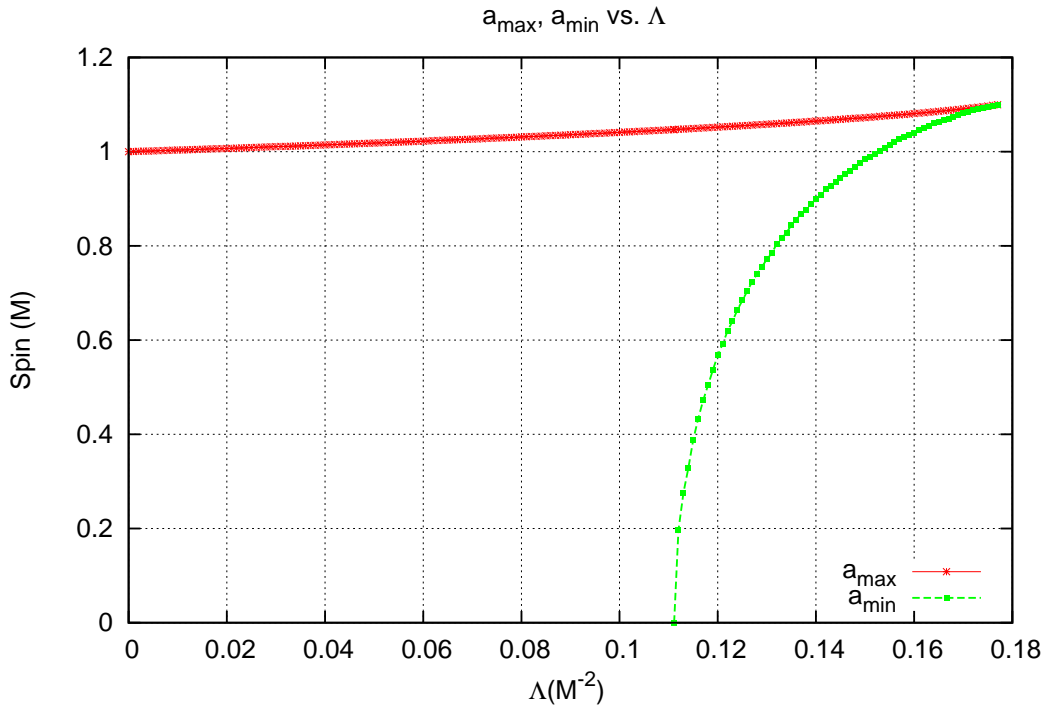}
\caption{Plot of $a_{max}$ and $a_{min} $ versus $\La$ with the
black hole mass $M = 1$. The upper, solid (red) curve ($\ast$
marks) represents $a_{max}$ for a KdS black hole. Note that
$a_{max} > M $ for all values of $\La
>0$ and $a_{max}=M$ only for the asymptotically flat Kerr case
($\La=0$). The lower, dotted (green) curve (square marks)
represents $a_{min}$, the minimum spin value the black hole needs
to have in order to preserve its horizon structure. $a_{min}$
becomes non-zero only for values of $\ti{\La}$ greater than $
1/9=0.111\overline{111} $ (the Schwarzschild-de Sitter limit). The
two curves meet at the maximum value of $\bar{\La} = 0.1778/M^2$.
For $\La > \bar{\La}$ there is no black hole, but only a naked
singularity. The ultimate extremal spin value at $\La=\bar{\La}$
equals $\bar{a} \approx 1.10084 M $.} \label{fig:KdS_a_vs_Lambda}
\end{figure*}
a conservative initial guess for the spin like $a=0.99$. Then we
increase $a$ until we meet the condition in
Eq.(\ref{eq:amax_cond}). The final value reached by the numerical
run is the maximal spin value $a_{max}$ that makes the Kerr black
hole extremal. For example, for $\ti{\La}=0.1$, the maximum value
for the spin is $\ti{a}_{max}=1.0409813 $ which is slightly above
the asymptotically flat Kerr bound $a=M$. This means that the
black hole can be ``overspun" yet still maintain its horizon
structure so long as $a < a_{max}$. We present the results of a
numerical evolution of $a_{max}$ as a function of $\La$ in Fig.
\ref{fig:KdS_a_vs_Lambda}. In the figure, the upper, solid (red)
curve represents $a_{max}$ for a KdS black hole of $M=1$ at a
given $\La$. Note that $a_{max} > M $ for all values of $\La >0$
and we only obtain $a_{max}=M$ for the pure Kerr case ($\La=0$).
Also in Fig. \ref{fig:KdS_a_vs_Lambda}, the lower, dotted (green)
curve represents $a_{min}$, the minimum spin value the black hole
needs to have in order to preserve its double horizon structure.
Notice that $a_{min}$ becomes non-zero only for values of
$\ti{\La}$ greater than $ 1/9=0.111\overline{111} $. The reader
will recall that 1/9 corresponds to the Schwarzschild-de Sitter
limit of $ \La < 1/ 9M^2$. So the green curve confirms what we
already know: One can not have a non-spinning ($a=0$) SdS black
hole for values of $\ti{\La}$ exceeding 1/9. {\it Thus, if we are
to have a black hole in a $\La
> 1/9M^2$ universe, it must be spinning}. This could have interesting
consequences on primordial black hole formation scenarios during
the inflationary era of the early Universe. If we assume that the
cosmology ($\Lambda$) is specified a priori, as it could be during
inflation, then we have a minimum spin requirement for the black
holes forming during this epoch. This could potentially forbid the
formation of Schwarzschild black holes.

Furthermore, as $\La$ increases, the spin lower bound $a_{min}$
increases as well, thus shrinking the parameter space of
admissible Kerr black holes in a de Sitter background. Once again,
there is a physical interpretation for this behavior. As is well
known, the Schwarzschild horizon is larger in size than the Kerr
horizon of the same mass, so as we consider larger black hole
spin, the horizon shrinks in size. As we consider larger $\La$,
the only black holes that can exist are those with smaller
horizons, i.e. larger spin values: thus the need for a Kerr black
hole to have a minimum spin value in a de Sitter background.
Furthermore, the two curves in Fig. \ref{fig:KdS_a_vs_Lambda} meet
at $\La= \bar{\La}\approx 0.1778 M^{-2} $, the maximum value of
$\La$ allowing a black hole. If the cosmological constant is any
larger than this ultimate value then the black hole can not hold
itself together against the cosmic repulsion and simply leaves its
place to a naked singularity, although how ``naked" this
singularity may be is open to debate. This is further discussed in
Section \ref{sec:Conc}. The extremal spin value $\bar{a}$ at
$\La=\bar{\La}$ is also the ultimate value for black hole spin:
$\bar{a}\approx 1.10084 M $.

It is straightforward to verify that $a_{max} $ represents an
extremal Kerr black hole. We do this by computing the surface
gravity of the KdS black hole at $r=r_+$. As computed explicitly
in the appendix \ref{sec:app2}, the surface gravity of the Kerr-de
Sitter black hole can be written as follows:
\be \kappa = \f{1-\f{\La}{3}r_+^2}{4Mr_+} \partial_r\Delta_r
|_{r=r_+} . \label{eq:kappa_1} \ee
(The expression given in \cite{GLPP} looks different, but can be
manipulated to exactly equal Eq.(\ref{eq:kappa_1})). Since by
definition an extreme black hole is one where the two horizons
(roots of $\Delta_r$) $r_-$ and $r_+$ meet, one automatically has
an extremum of $\Delta_r$ at that meeting point i.e. $
\partial_r\Delta_r |_{r=r_-=r_+} = 0 $. Thus, the
KdS black hole with the doubly degenerate horizon ($r_-=r_+$) is
indeed an extremal black hole.



\section{Global Structure} \label{sec:Globe}
\begin{figure*}
\includegraphics[height=8cm]{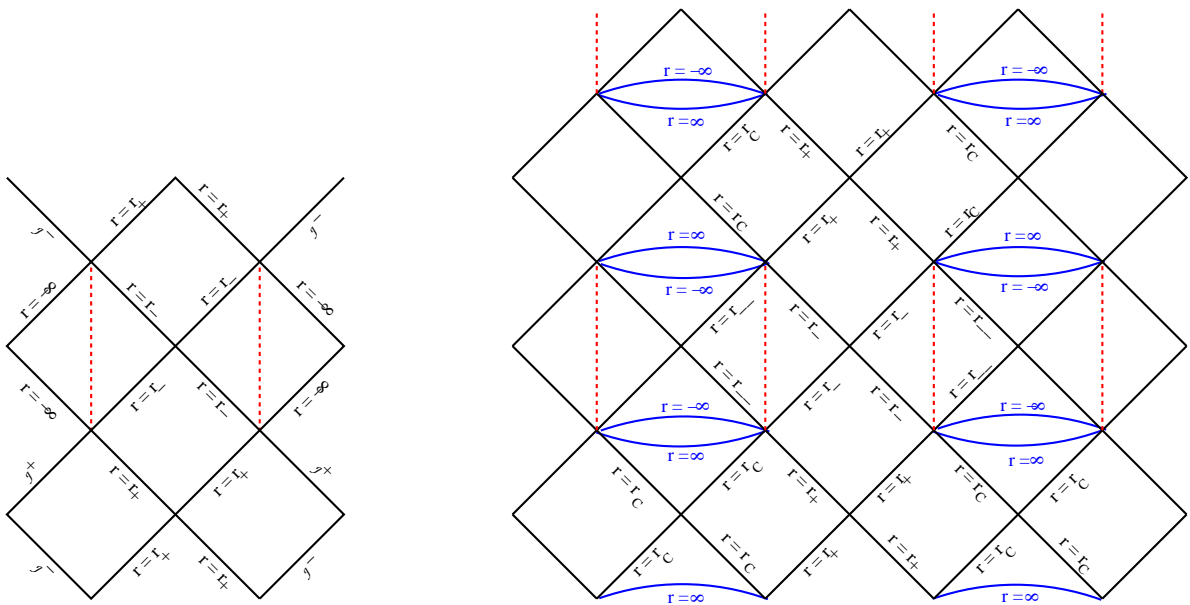}
\caption{The Penrose-Carter diagrams for Kerr (left) and Kerr-de
Sitter (right) spacetimes along the $\theta=0$ cut \cite{GH}. When
viewed together, the differences between the two spacetimes become
obvious. In KdS there are four horizons labelled $r_{--},r_-, r_+$
and $r_C$. As in Kerr, $r_\pm$ are the black hole horizons, but
there are two extra horizons, $r_{--}, r_C$ which are cosmological
in nature. The infinities at $r=\pm \infty$ are disjoined. The
dashed (red) vertical lines represent the curvature singularities
at $r=0$. These are actually ring singularities that lie on the
$\theta=\pi/2 $ plane. The Kerr singularity is naked when viewed
from the region $r<0$ whereas in KdS, the same singularity is
cloaked behind the horizon $r_{--} $ for observers at $ -\infty <
r < r_{--} < 0 $. The conformal diagram for KdS continues ad
infinitum in all directions. As always, the transverse directions
($\theta,\phi$) have been suppressed.}\label{fig:KdS_penrose}
\end{figure*}

We start by presenting the Penrose diagram for the Kerr-de Sitter
spacetime along the positive z-axis (symmetry axis $\theta=0$) in
Fig.\ref {fig:KdS_penrose}. Although the Penrose diagrams for Kerr
spacetime are usually presented along an equatorial slice
($\theta=\pi/2$), our choice for the spatial cross-section
resembles the spherically symmetric Reissner-Nordstr\"{o}m-de
Sitter more and is also the choice of Hawking \& Gibbons in
\cite{GH}. In fact, the Penrose diagram for KdS looks similar to
that of Kerr (cf. \cite{MTW}, \cite{Wa}) except for the extra
horizons at $r=r_{--}, r_C$ and the fact that conformal infinity,
$\mathcal{J}^+ $, is replaced by a spacelike surface. This is
consistent with the spacelike $\mathcal{J}^+ $ also found in de
Sitter spacetime ($\Lambda>0$). Just as in Kerr, one has the black
hole event horizon at $r=r_+$, the inner black hole [Cauchy]
horizon at $r=r_-$ and a ring singularity at $r=0$ which is
pictured as the red dashed lines in Fig. \ref{fig:KdS_penrose}.
But unlike in Kerr, the ring singularity is no longer naked in the
region where $r<0$. It is cloaked beyond yet another horizon
$r_{--} <0 $ and past this horizon, there is yet another infinity
$r=-\infty$. A few words of caution about this diagram: {\bf (1)}
The infinities at $r=\infty$ and $r=-\infty$ are completely
disjoined, {\bf (2)} Technically speaking, the ring singularity
lies on the $\theta=\pi/2$ plane and not on the $\theta=0$ cut we
are taking, but this somewhat misleading graphical representation
of the singularity is standard treatment in the literature whether
for Kerr or Kerr-de Sitter spacetimes (cf. \cite{GH}, \cite{MTW}).

One can identify different regions of this spacetime. This is
indeed what we will do explicitly in this section, first for
Reissner-Nordstr\"{o}m-de Sitter spacetime then for KdS. The
identifications will help reveal the physical picture for these
spacetimes: the spatial manifold is a 3-sphere that contains a
black/white hole at its antipodal points (poles at $\chi=0, \pi$,
see Eq.(\ref{eq:dSmetricTchi}) ). Here and henceforth, when we say
that a black hole (a 4-dim. object) is contained within a
3-sphere, we mean that its 2-dim. event (or apparent) horizon is
centered at the pole of the 3-sphere.

Before continuing with the interpretation of the Kerr-de Sitter
Penrose-Carter diagram, we first discuss the
Reissner-Nordstr\"{o}m-de Sitter (RNdS) (\cite{DWDW}) and
Kastor-Traschen (KT) (\cite{KT}) solutions. Just as one heavily
draws on Reissner-Nordstr\"{o}m spacetime to understand the
conformal structure of Kerr spacetime, here we will draw on RNdS
spacetime to suggest an interpretation of KdS.
We will find that a section of the extended KdS manifold can be
interpreted as a 3-sphere containing two counter-rotating Kerr
holes: a black hole and a white hole located at antipodal points.
A similar interpretation was first mentioned in \cite{MM} for the
RNdS case; it can be shown in a straightforward manner for the
case of ``extremal" (charge Q equaling mass M) RNdS spacetime. It
turns out that this extremal RNdS spacetime is a single-black-hole
Kastor-Traschen spacetime; there is an exact coordinate
transformation relating the two solutions. (By construction, the
KT solution can accommodate as many black holes as one wishes.)
Most of the work on this was done by Brill and collaborators in
\cite{BH}, \cite{BHKT}, \cite{Br} and throughout this article, we
rely on their results repeatedly.

Here, we are mostly interested in the single black hole KT
spacetime and its coordinate charts. Nevertheless, we include a
brief section on the many black hole KT solution in appendix
\ref{sec:app3} for the sake of completeness.


\subsection{Reissner-Nordstr\"{o}m-de Sitter
Spacetime}\label{ssec:RNdS} The spherically symmetric, three
parameter ($ M, \Lambda, Q $) solution to Einstein's equation was
also first written down by Carter in \cite{DWDW}. Here $Q$
represents the electric charge of the black/white hole. The line element
for this solution in static coordinates is
\be ds^2 = -f dt^2 + f^{-1} dr^2 + r^2 d\Omega_2^2,
\label{eq:RNdS_metric} \ee
where
\be f = f(r) = 1-\f{2M}{r} + \f{Q^2}{r^2} - \f{\Lambda}{3} r^2,
\label{eq:RNdS_f} \ee
and $d\Omega_2^2 = d\theta^2 + \sin^2\theta d\phi^2 $ is the
metric on the 2-sphere ($S^2$). Once again, the location of the
horizons is determined by the roots of the quartic equation $ f =
0 $. As before, only for certain values of $\Lambda <
\Lambda_{max}$, does one get three real positive roots $r_- < r_+
< r_C$ representing the inner, outer black hole and the de Sitter
horizons. This is what we will refer to as the generic case. There
are, of course, non-generic cases where the quartic has only one
real positive root or degenerate double or triple roots. These
more eccentric cases are mentioned in detail in \cite{BH},
\cite{BHKT}, \cite{Br}. Here, we are concerned with the generic,
three positive root case. The Penrose diagram for this case looks
exactly like the Penrose diagram displayed for KdS universe in
Fig. \ref{fig:KdS_penrose}.

For the generic case, where $f>0$ and $ r_+ < r < r_C $, the
$t=constant$ surfaces are free of singularities. The spatial
geometry on such a $t=constant$ hypersurface is given by its
spatial 3-metric
\be ds_3^2 = f^{-1} dr^2 + r^2 d\Omega_2^2 \label{eq:RNdS_3metric}
. \ee
This can be embedded in 4-dimensional Euclidean space $ ds_E^2 =
dW^2 + dr^2 + r^2 d\Omega_2^2 $ by the following definition
\cite{BH}
\be W \equiv \int \sqrt{f^{-1}-1}\; dr . \label{eq:W_embed} \ee
%
The embedding hypersurface has a minimum radius at the throat of
the black hole ($r=r_+$) and maximum radius at the cosmological
horizon ($r=r_C$). The spatial topology is $ S^2 \times R^1 $.
At this point, let us mention that removing both poles of a
3-sphere changes its topology from $S^3$ to $S^2 \times R^1 $,
which is the topology of spatial sections of RNdS spacetime.
Certainly, hiding the poles of the 3-sphere behind black/white
hole event horizons is a good way to change the topology to the
desired one. This is in line with Mellor and Moss's (\cite{MM})
physical interpretation of the extended RNdS spacetime as
containing black/white holes at the poles of the de Sitter
3-sphere. In much of the literature the statement is made that
there are two {\it black} holes (more precisely apparently no
distinction is made between black holes and white holes). We will
show that different spacelike slices in the same domain can cut a
future horizon (BH) and/or a past horizon (white hole). In
\cite{Br}, Brill interprets these event horizons as wormhole
mouths at the antipodal points of the large de Sitter universe.


%

To further clarify Mellor and Moss's interpretation \cite{MM},
consider the RNdS spacetime given by $ |Q| = M $.
As in the $a=M$ case in KdS, $|Q|=M$ in RNdS with $\La>0$ is not
extremal in the sense of yielding zero surface gravity for the
black hole (i.e. $r_+=r_-$). For $0<M^2\Lambda < 3/16 $, the
spacetime has three horizons. At $M= M_{crit} \equiv
(3/16\Lambda)^{1/2}$, the outer event horizon and the cosmological
horizon merge, thus yielding a degenerate cosmological horizon
accompanying the inner Cauchy horizon at $r=r_-$ (see eg. Fig 1 of
\cite{HNS}). For $M>M_{crit} $, the spacetime has only the
cosmological horizon and naked singularities (see \cite{BH},
\cite{BHKT}, \cite{Br} for details and Penrose diagrams of these
separate cases). Such solutions thus present potential violations
of the cosmic censorship conjecture of Penrose \cite{Pe}. The
issues of cosmic censorship violation are not the main focus of
this article so we postpone our discussion to the end of section
\ref{sec:Conc}. More details can be found in \cite{BHKT},
\cite{Br} and \cite{Wa2}.

In static coordinates the RNdS metric for
the $|Q|=M$ case looks like
\be ds^2 = -f dt^2 + f^{-1} dr^2 + r^2 d\Omega_2^2,
\label{eq:RNdS_ext_metric} \ee
where we now have
\be f = \left(1-\f{M}{r}\right)^2 - \f{\La}{3}r^2
\label{eq:RNdS_f_ext} . \ee
As before, the horizons are located at $f=0$ and there is a
curvature singularity at $r=0$.

Because static coordinates break down at the horizons
\footnote{This breakdown is well known in the Schwarzschild
spacetime expressed in Schwarzschild coordinates.}, it is more
useful to look at the RNdS spacetime with another set of
coordinates (chart). Such a chart is given by \emph{cosmological}
coordinates that smoothly cover the entire region from $r=0$ to
$r=\infty$. A version of various charts covering different regions
of de Sitter spacetime is presented in detail in \cite{HE}.

The $|Q|=M$ RNdS metric can be written in terms of these cosmological
coordinates \cite{KT}. The metric in the new coordinates is given by
\be ds^2= -\f{d\tau^2}{U^2} + U^2 (dR^2 + R^2 d\Omega_2^2),
\label{eq:RNdS_met_KTcoord} \ee
where
\be U =  H \tau + \f{M}{R} \label{eq:KT_U} \ee
in which $ H \equiv \pm (\Lambda/3)^{1/2} $ is the Hubble
constant. The metric above uses the $H>0$ value. The coordinate
transformation equations from the static coordinates
($t,r,\theta,\phi$) to the cosmological coordinates ($\tau,
R,\theta,\phi $) are
\ba r & = &  H \tau R + M ,\\
   t & = & \f{1}{H} \ln H\tau - h(r), \\
   \f{dh}{dr} & = & -\f{H r^2}{(r-M) f}. \ea
It is straightforward to verify that one can go from
Eq.(\ref{eq:RNdS_ext_metric}) to Eq.(\ref{eq:RNdS_met_KTcoord})
using this coordinate transformation. The curvature singularity is
located at $U=0$ i.e. $ H\tau R = -M $. The event horizon $r_+$ is
at $(\tau, R) = (\infty, 0) $, the cosmological horizon $r_C$ is
at $(\tau,R) = (0, \infty)$ and the inner horizon $r_-$ is at
$(\tau,R) = (-\infty,0)$. The details of how these values come
about are explained thoroughly in \cite{NSH}. In fact, one could
do the same coordinate transformation from static to cosmological
coordinates for the regular de Sitter spacetime. This is done in
\cite{Br} explicitly and the procedure parallels  the one we have
outlined above.

\begin{figure*}
\includegraphics[height=8cm]{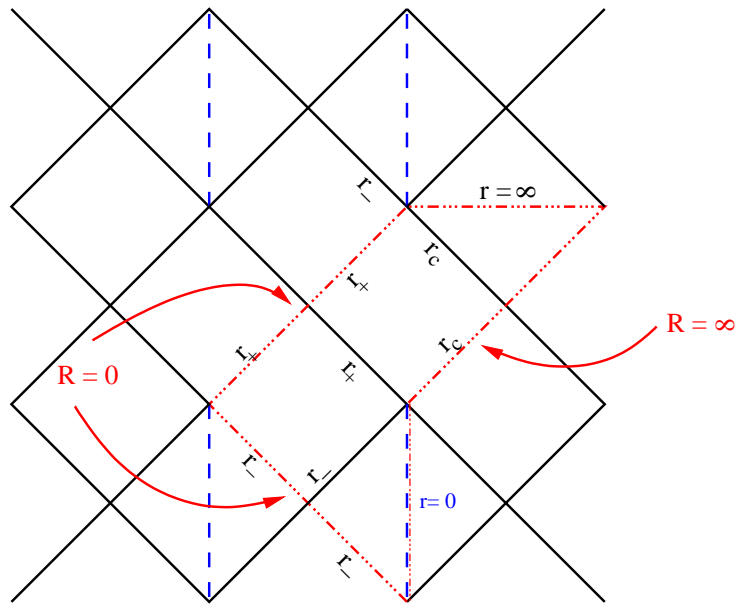}
\caption{Penrose diagram of the extended $ |Q|=M $
Reissner-Nordstr\"{o}m-de Sitter spacetime. The region bounded by
the 3-dotted-dashed (red) line represents a single ($\tau,R$)
chart. $r$ is the radial coordinate in the old static coordinates
whereas $R$ is in the new cosmological coordinates. For the $|Q|
\ne M$ generic case, a single $\tau$ chart is not sufficient to
cover the region enclosed by the red line segments
(dashed-triple-dotted). Figure 5 of \cite{BH} illustrates the
generic case for this coordinate chart.} \label{fig:TRcharts}
\end{figure*}

One advantage of switching to cosmological coordinates is that a
single ($\tau, R$) chart now covers four ($t,r$) charts ranging
from $r= 0$ to $r=\infty$ and encompassing all three horizons. The
boundaries of this chart are the singularity, the past inner
horizon $r_-$, the future event horizon $r_+$, the past
cosmological horizon $r_C$ and the future conformal infinity at
$r=+\infty$. Fig. \ref{fig:TRcharts} illustrates the region that a
single ($\tau,R$) chart covers against the backdrop of the
extended RNdS spacetime. It should be noted that for the generic
$|Q| \le M$ case, a single $\tau$ chart is not sufficient to cover
the region shown in Fig. \ref{fig:TRcharts}. A `lense-shaped
region' \cite{BH} in between $R_\pm = M \pm \sqrt{M^2-Q^2} $ would
remain {\it uncovered} by the cosmological coordinates. More on
this can be found in \cite{BH} and \cite{Br}.

A few words should be mentioned about the nature of the
singularity in this coordinate system. Since $H, M>0$ and the
singularity is located at $U = H\tau +M/R = 0 $, it exists only
for $\tau<0$. That is, there is no curvature singularity for $\tau
> 0$; for positive values of $\tau$, the $\tau=constant$
hypersurfaces are regular. These non-singular surfaces are
asymptotically flat as $R\rightarrow \infty$ and they have a
throat-like cylindrical geometry near $R=0$, which can be seen by
considering the metric of Eq.(\ref{eq:RNdS_met_KTcoord}) at some
$\tau=constant$ slice, near $R=0$. This 3-dimensional line element
reads
\be dl^2 = \f{M^2}{R^2} dR^2 + M^2 d\Omega_2^2.
\label{eq:3cyl_met} \ee
This is the metric of an infinitely long throat with
cross-sectional area $4\pi M^2$. At $\tau=0$, the spatial metric
is still regular and looks like Eq.(\ref{eq:3cyl_met}) everywhere.
As $\tau$ is rewound to negative values starting with $0^{-}$, a
singularity forms at $R=\infty $ and as $\tau \rightarrow
-\infty$, the singularity moves in closer with ever decreasing
values of $R$ and truncates the 3-cylinder more and more. The
spacetime shrinks to a point at $\tau = -\infty$. Instead of going
backward in time, if we simply fast-forward, we see that what we
have is a spacetime that expands out of a single singular point.
Because of this property, the ($\tau,R$) chart with $H>0$ is
called the \emph{expanding} chart and usually is labelled by
($\tau_+,R $). From a physical perspective, this agrees with our
choice of the positive root for $H=\pm (\Lambda/3)^{1/2} $.

\begin{figure*}
\includegraphics[height=8cm]{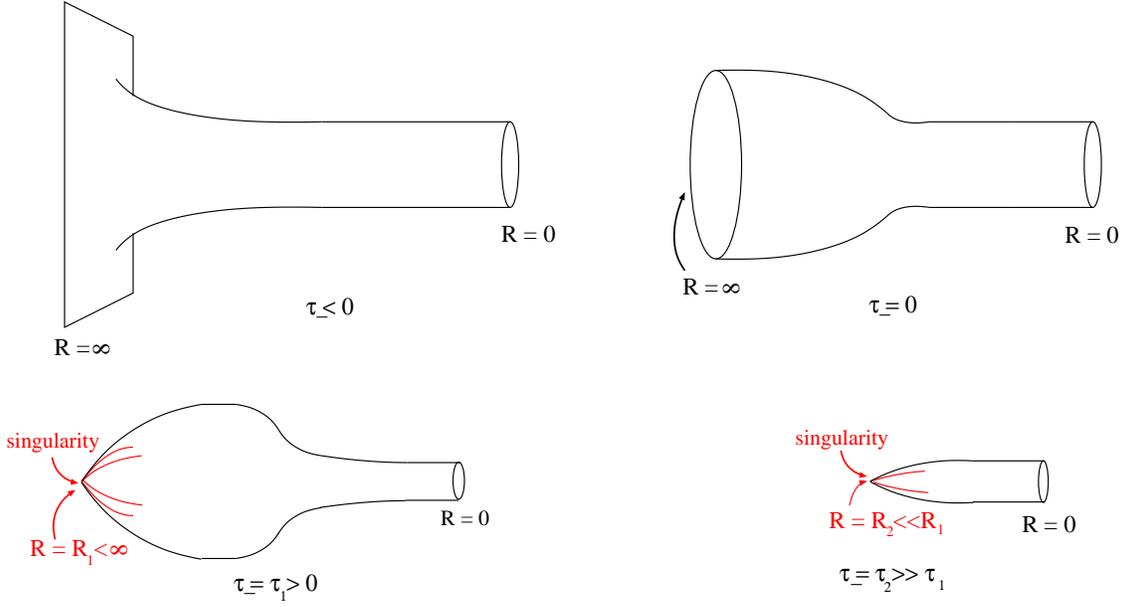}
\caption{Time evolution of a contracting ($\tau_-,R $) chart
\cite{BHKT}. The singularity forms at $(\tau_-,R) = (0, \infty) $
and grinds the space down as $\tau_-$ increases tending to $R=0$
at $\tau_-\rightarrow \infty $. The Penrose diagram corresponding
to this process is presented in Fig. \ref{fig:RNdS2charts}. The
contracting chart is the region to the left of the blue dashed
line. As can be seen from the figure, at early times ($\tau_-<0$),
there is no singularity, which then forms at $\tau_-=0$ and
proceeds to grind the spacetime down for $\tau_->0$ until there is
nothing left but the Big Crunch itself.} \label{fig:KT_time_evol}
\end{figure*}

There is a {\it companion} chart to the expanding chart that
covers another equal size section of the RNdS spacetime. This
chart is called the \emph{contracting} chart and labelled by ($
\tau_-,R  $). The metric of Eq.(\ref{eq:RNdS_met_KTcoord}) now is
written with
\be U = H \tau_- + \f{M}{R}= -|H|\tau_- + \f{M}{R}, \ee
with the negative root of $H$ chosen. Because $\tau$ enters the
metric Eq.(\ref{eq:RNdS_met_KTcoord})  only in the combination
$H\tau$ the ($\tau_-,R$) chart is actually the time reversed
version of the expanding chart. The threshold moment is again
$\tau_-=0$, but this time the singularity forms at $R=\infty$ as
$\tau_-$ becomes positive since at $U=0$, one has $ |H|\tau_- =
M/R $. This obviously only holds true for positive values for
$\tau_- $. As $\tau_-$ increases, the singularity approaches `in'
from $R=\infty$ and the entire spacetime collapses to a
singularity at $R=0$ as $\tau_- \rightarrow \infty $ (See Fig.
\ref{fig:KT_time_evol} \cite{BHKT} for an illustration of how the
spatial hypersurfaces evolve). Clearly, things happen in the
$\tau_-$ chart in the exact opposite order of the way they happen
in the $\tau_+$ chart.
A similar version of
these expanding/contracting cosmological charts were constructed for de
Sitter spacetime by Brill in \cite{Br}. Of course there, one
 has no curvature singularity at $U=0$.

\begin{figure*}
\includegraphics[height=8cm]{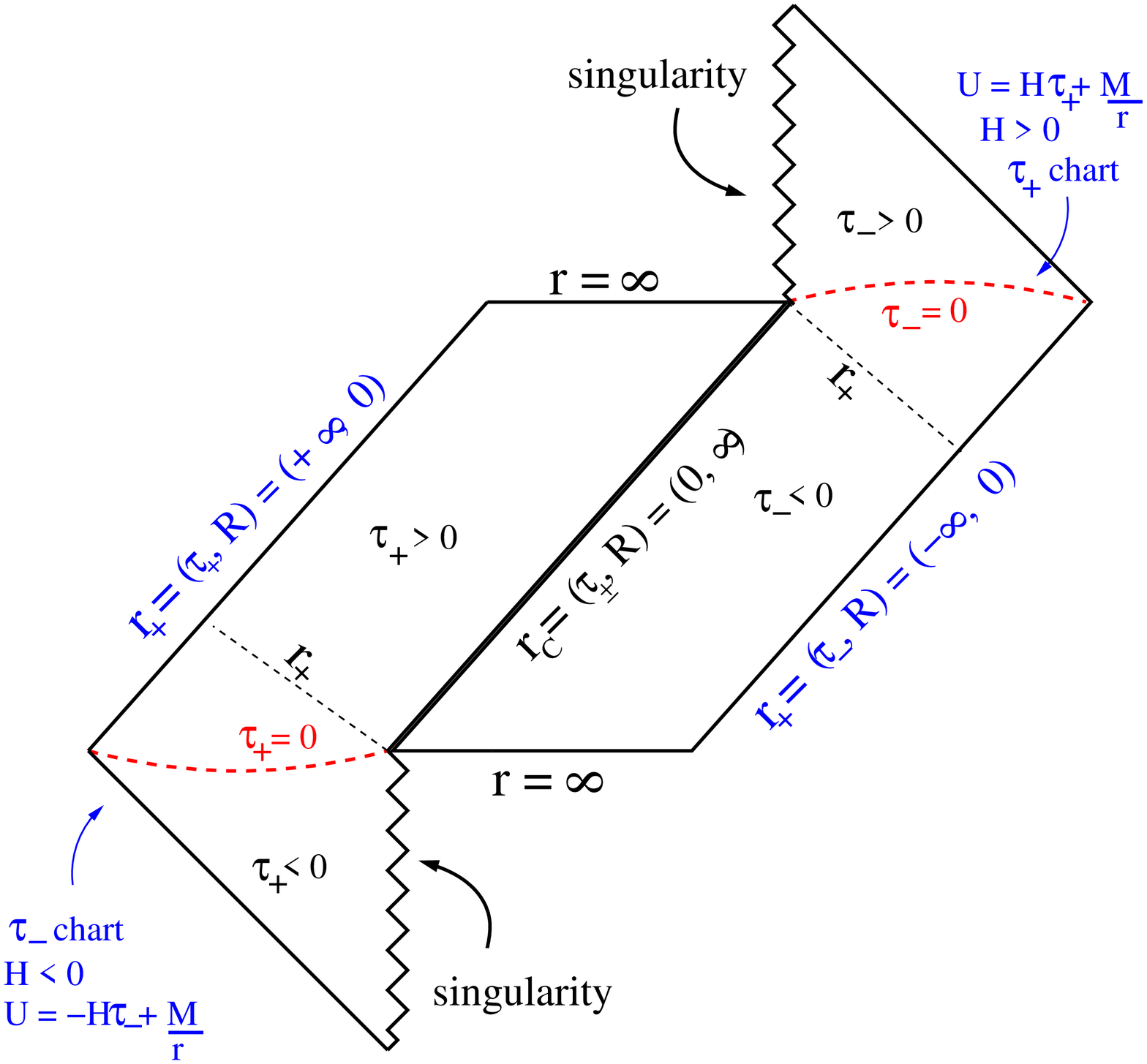}
\caption{The Penrose diagram showing the $\tau_\pm$ charts
covering regions of the extended $|Q|=M$ RNdS spacetime. The
left-hand $+$ region represents part of the spacetime that expands
out of a singularity and eventually becomes regular everywhere.
The right-hand ``$-$'' region contains part of the spacetime that
eventually shrinks to singularity. The fainter, dotted $45^\circ$
lines also represent black hole horizons $ r_+$ that the
cosmological coordinates cross without breaking down.}
\label{fig:RNdS2charts}
\end{figure*}

\begin{figure*}
\includegraphics[height=6cm]{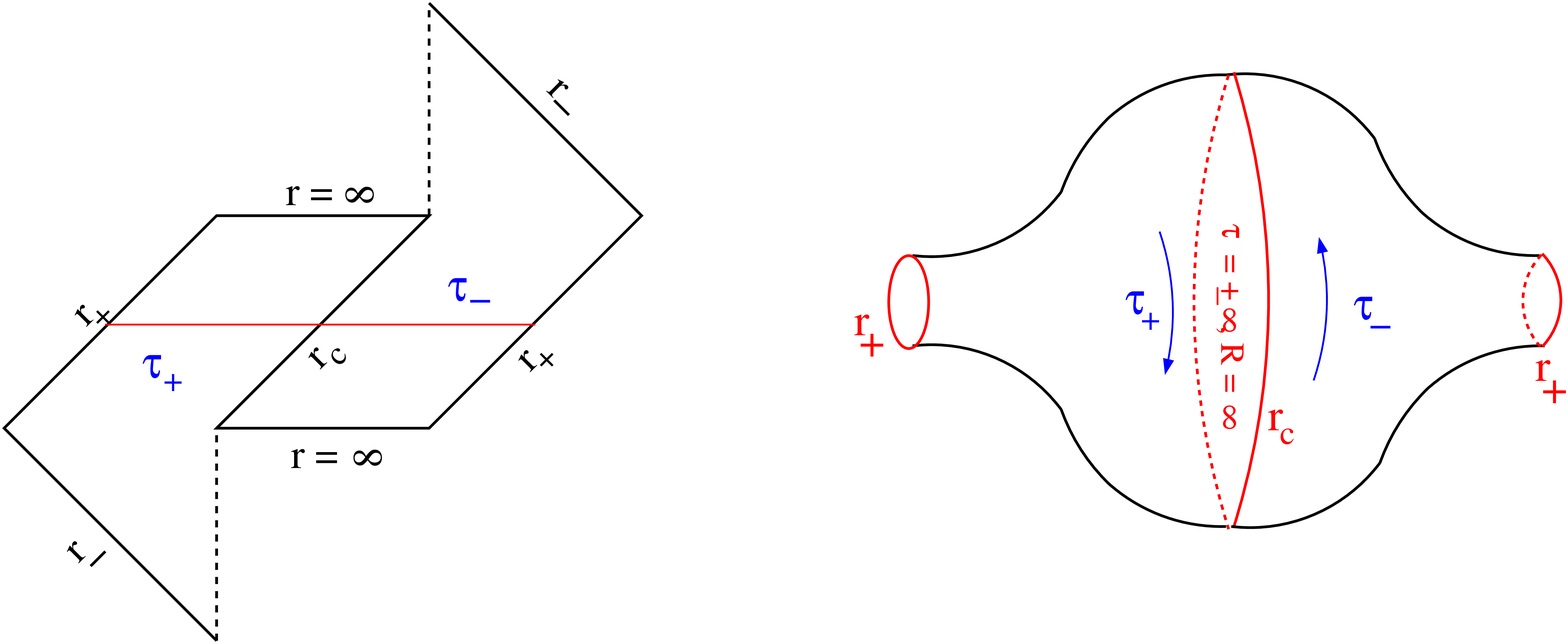}
\caption{A simplified Penrose diagram of the $\pm$ charts of Fig.
\ref{fig:RNdS2charts} along with the corresponding schematic
embedding diagram along the red (solid line) hypersurface, which
is a time-symmetric cut of the spacetime. Both charts have the
same metric at $r=r_C$, thus they are joined together there. The
corresponding embedding diagram is also glued at $r=r_C$, which
gives us the $S^2\times R^1 $ topology since the poles of the
3-sphere are behind event horizons. Non-time-symmetric cuts would
result in a similar 3-geometry with differing sizes for the `bead'
and the `throats'. Such a non-time-symmetric cut is shown for
Schwarzschild-de Sitter on the upper-right hand panel of Fig.
\ref{fig:SdS} where the solid (red) curve represents the
3-geometry of a time-symmetric slice and the dashed (green) curve
of the non-time-symmetric case. The time reversing between the
charts is shown by the blue arrows. For RNdS this implies a
reversing of the direction of {\bf E}-field lines; for Kerr-de
Sitter, it would imply counter-rotation of two holes located at
the poles.} \label{fig:KdS_identify}
\end{figure*}

The $\tau_\pm$ charts together cover a large portion of the
extended RNdS spacetime. Fig. \ref{fig:RNdS2charts} displays both
of these charts in the Penrose diagram of RNdS. The black hole
event horizon $r_+$ is given by ($\tau_\pm,R$) = ($\pm \infty, 0
$). The cosmological horizon $r_C$ is located at ($\tau_\pm,R$) =
($0,\infty$). Since both $\pm$metrics give $U= M/R \rightarrow 0 $
at $r=r_C$ (i.e. $\tau_\pm = 0 $) , they can be isometrically
identified, i.e. glued together. This is shown more clearly in
Fig. \ref{fig:KdS_identify}. Furthermore, this identification i.e.
gluing a $+$chart to a $-$chart repeats itself ad infinitum in the
horizontal direction.

\begin{figure*}
\includegraphics[height=6cm]{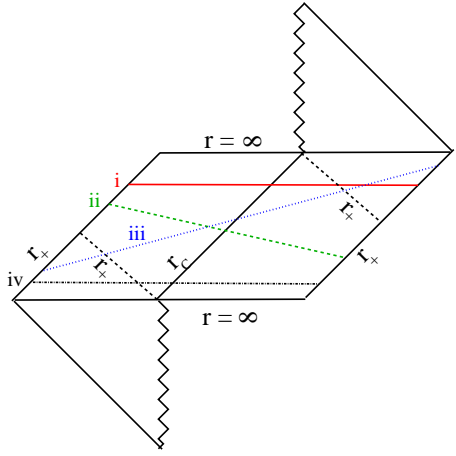}
\caption{Very similar to the slicing of SdS spacetime shown in
Fig. \ref{fig:SdS}. We use the same color coding and line types as
before: Red, solid (i); green, dashed (ii); blue, dotted (iii);
black, dash-triple-dotted (iv). These slices intersect BH/WH at
$45^\circ $ lines marked $r_+$ (either solid or dotted). These
slicings respectively give $ BH-BH, BH-WH, WH-BH, WH-WH $ pairings
within the chosen hypersurface.}\label{fig:KdS_hypersurfaces}
\end{figure*}

It can be seen from Fig. \ref{fig:RNdS2charts} that timelike lines
to the ``left'' of the stitching boundary $r_C$ must all go
through the left-hand boundary at $r=r_+$, whereas timelike lines
in the region to the right of $r_C$ all emanate from the
right-hand $r_+$.
We find that in the literature, the objects beyond the two
horizons at $r=r_+$ are simply referred to as black holes, but in
reality the question of which kind of hole one has depends on the
type of spacelike hypersurface one is in. This is illustrated in
detail in Fig. \ref{fig:KdS_hypersurfaces}. We see that depending
on the hypersurface chosen, the global geometry can be given by
any one of the four $BH-BH, WH-BH, BH-WH, WH-WH $ combinations.
Regardless of their nature, these holes are located at the poles
of the de Sitter 3-sphere. When patched together in this way there
is a global time coordinate $\tau$ which covers most of the pasted
patch, and which increases ``upward" in the resulting pasted
spacetime, even though $\tau_-$ decreases as $\tau$ (and $\tau_+$)
increase. Because the $\tau_-$ patch is identical to the $\tau_+$
patch evolving oppositely in terms of a global observer, to such
an observer there is a (formal) time reversal between them, so the
electric fields (hence the electric charges) are reversed. (The
components of the Maxwell tensor transform as $F^{0i} \rightarrow
- F^{0i}$.) Electric field lines leaving one black hole will enter
the other one at the opposing antipodal point. This is precisely
Wheeler's ``charge without charge". If desired, one could further
identify the black hole throats (horizons) with each other and
turn the $S^2\times R^1 $ topology into a $S^2 \times S^1 $
topology. This identification would transform the portion of the
RNdS spacetime covered by the ($\tau_\pm,R$) charts to a Wheeler
wormhole introduced in \cite{W}. This is the picture presented by
Brill in \cite{Br}.

\subsection{Kerr-de Sitter from Reissner-Nordstr\"{o}m-de Sitter}
We will now demonstrate that the extended KdS manifold can be
viewed as a 3-sphere containing a black hole at one pole and
another at the opposite pole. In section \ref{ssec:RNdS}, we
showed this concretely for the RNdS spacetime. Here, we will show
that the same interpretation holds for KdS. Although this may be
obvious to some (especially since Reissner-Nordstr\"{o}m and Kerr
share the same causal structure), we will still take the time to
make the connection more clear. Our results to this end will not
be analytically exact, but are based on a simple distortion of
RNdS spacetime in which the KdS spacetime outside the event
horizons remains free of singularities and other kinks. Our
specific example is the very restricted case of $a=M$ Kerr-de
Sitter, but we argue that it is representative of the larger
family of KdS solutions.

Ideally, we would like to find an exact coordinate transformation
that would turn the KdS metric to a single black hole KT metric of
Eq.(\ref{eq:RNdS_met_KTcoord}). The KdS metric of
Eq.(\ref{eq:KdS_metric1}) is rather complicated, so finding such a
coordinate transformation has thus far been elusive. Instead, we
make the KdS metric resemble the RNdS metric of
Eq.(\ref{eq:RNdS_met_KTcoord}) and use the already available
coordinate transformation introduced in section \ref{ssec:RNdS}.
To this end, we first take a $\:\theta=0$ polar cut of the metric
in Eq.(\ref{eq:KdS_metric1}) (due to axisymmetry $\phi$ can be set
to equal any value from 0 to $2\pi$). Then, in analogy to the
$|Q|=M$ RNdS case, we focus our attention solely on the $a=M$ KdS
solution. We do this so that we can rewrite $\Delta_r$ of
Eq.(\ref{eq:Delta_r}) as follows
\ba \Delta_r &=&  r^2-2Mr+a^2-\f{\La r^2}{3}(r^2+a^2) , \nn \\
 & = & (r-a)^2 -\f{\La r^2}{3}(r^2+a^2). \label{eq:new_Delta_r}
 \ea
Furthermore, we define a new time coordinate $T$ by
\be T \equiv \f{t}{1+\f{\La}{3}a^2}. 
\label{eq:new_T} \ee
This is simply a rescaling of the time coordinate by a constant
factor. Putting all the new changes together into our $a=M$ KdS metric, we obtain
\be ds^2  = -\left[\left(\f{r-a}{\sqrt{r^2+a^2}}\right)^2 - \f{\La
r^2}{3}\right] dT^2 + (-g_{TT})^{-1} dr^2 ,\label{eq:KdS_2metric2}
\ee
where $g_{TT}$ equals the expression inside the large square
brackets. This looks very similar to the extremal RNdS metric
given by Eq.(\ref{eq:RNdS_ext_metric}), which we display one more
time (with $a$ replacing $M$ for the sake of making the comparison
easier to see):
\be ds^2 = -\left[\left(1-\f{a}{r}\right)^2 - \f{\La r^2}{3}
\right] dt^2 + (-g_{tt})^{-1} dr^2 \label{eq:RNdS_again}. \ee
We know that Eq.(\ref{eq:RNdS_again}) corresponds exactly to the
single mass Kastor-Traschen solution which can be covered by two
charts ($\tau_\pm, R $). We have also shown that the part of the
extended RNdS covered by these two KT charts can be physically
interpreted to be the de Sitter 3-sphere with black/white holes at
its poles. The extremal KdS metric of Eq.(\ref{eq:KdS_2metric2})
is very nearly identical in form to the extremal RNdS of
Eq.(\ref{eq:RNdS_again}). Let us see if we can massage the KdS
metric (\ref{eq:KdS_2metric2}) further to make it look more like
Eq.(\ref{eq:RNdS_again}). We start by defining a new coordinate
$\bar{r}$ in $a=M$ KdS spacetime by the following:
\be \b{r}^2 \equiv r^2+a^2 \rightarrow \b{r}d\b{r} = r\:dr .\ee
%
Since the maximum value of $a$ is only slightly above unity (see
section \ref{sec:KdS}), we have $a^2/M^2 \sim \mathcal{O}(1) $.
For $r/M>>1$, $ r \approx \b{r} $. The new coordinate substitution
also yields the following expressions:
\ba dr^2 & = & \f{\b{r}^2}{r^2}d\b{r}^2 = \f{\b{r}^2}{\b{r}^2-a^2}
d\b{r}^2 \equiv (1+\delta_1) d\b{r}^2, \\
\left(\f{r-a}{\sqrt{r^2+a^2}}\right) & = &
\left(\f{\sqrt{\b{r}^2-a^2}}{\b{r}}-\f{a}{\b{r}}\right)\equiv
\left((1-\delta_2) - \f{a}{\b{r}}\right), \ea
where
\ba \delta_1 &=& \f{a^2}{\b{r}^2-a^2}, \label{eq:delta1} \\
\delta_2 &=& 1-\f{\sqrt{\b{r}^2-a^2}}{\b{r}} . \label{eq:delta2} \ea
Above, instead of Taylor expanding in powers of $\b{r}$ we have
simply gathered all the troublesome terms under the `deviation'
terms $\delta_1$ and $\delta_2$.
Let us first redisplay the KdS metric of
Eq.(\ref{eq:KdS_2metric2}) with these new substitutions:
\be ds^2=-\left[\left((1-\delta_2)-\f{a}{\b{r}}\right)^2
-\f{\La}{3} \left(\b{r}^2- a^2\right)\right] dT^2 +
\f{(1+\delta_1)}{(-g_{TT})} d\b{r}^2 . \label{eq:KdS2metric3} \ee
Recalling our dimensionless tilde notation, we have
$\ti{a}=\ti{a}^2=1$. Because of this, the deviations $\delta_1$
and $\delta_2$ only grow to significant values near $\ti{\b{r}}
=1$. For example, even for $\ti{\b{r}} = 2$, the deviations do not
exceed 30\%. Furthermore, as $a=M$ is not the maximum value for
the spin of a Kerr black
hole in de Sitter background 
($\ti{a}_{max} \approx 1.1 $), the actual outer black hole horizon
$\b{r}_+ (\:= r_+^2+a^2)$ lies outside $ \b{r} = M $. Since the
conformal domain of interest is between the outer horizon $r_+$
($\b{r}_+$) and the cosmological horizon $r_C$ ($\b{r}_C$) having
a value for the spin less than the maximum ($a = M < a_{max}$)
guarantees that we are safe from the $\b{r} = a$ pole in
$\delta_1$ shown in Eq.(\ref{eq:delta1}). For example, for value
of $\ti{\La} = 0.001$ with $\ti{a}=1$ we have
\be \ti{r}_{+} \approx 1.02 \ \rightarrow \ \ti{\b{r}}_{+} \approx
1.5. \ee
Both of these are larger than 1. And obviously $\ti{r}_{+} = 1$
only at the $\La=0$ limit. Even with such a small value (close to
0) for $\La$ the deviations do not exceed order unity; e.g. for
$\ti{\b{r}}_{+} =1.5$ one has $\delta_1 = 0.8 $  and $\delta_2 =
0.25$.

Although one could argue that the deviation factors are large
enough to be significant near the black hole horizon ($\delta_1
\sim \ti{r}_+ \sim \mathcal{O}(1)$), this does not alter the
conformal structure of the spacetime in a serious manner. The
deviations are contributions to the metric only near the black
hole horizon (both $\delta_1$ and $\delta_2 \rightarrow 0 $ as
$\b{r} \rightarrow \infty $). As such, they only change
measurement of lengths in the vicinity of the event horizons.
However, rescaling lengths by at most a factor of $\sim$ 1.8 (for
$ \ti{\b{r_+}} \approx 1.5 $, $1+\delta_1 \approx 1.8$) is not a
significant change to the global structure of a given spacetime
manifold. As long as we do not poke any new holes in the manifold,
the general conformal picture should not change. This is the basis
for our argument that the general picture of the global structure
of Kerr-de Sitter should look like that of
Reissner-Nordstr\"{o}m-de Sitter. In short, even though KdS metric
can not be made to look exactly like the KT metric of
Eq.(\ref{eq:RNdS_met_KTcoord}), the 3-dimensional $\theta=0$
sub-spacetime of $a=M$ KdS looks very much like the $Q=M$ RNdS
metric. This resemblance is exact in the large $r$ limit. Near the
black hole horizons, the lengths are rescaled by nearly a factor
of two at most, but this is not a significant change to the global
structure of the KdS manifold.

As one has two oppositely charged anti-podal black/white holes in
the extended RNdS spacetime, here we expect to have
counter-rotating holes at the anti-podal points. The $\tau_-$
chart runs backward with respect to the $\tau_+$ chart (see Fig.
\ref{fig:RNdS2charts}) so one hole rotates oppositely to the
other, therefore yielding zero net angular momentum for the
extended solution. The right hand figure in Fig.
\ref{fig:KdS_identify} illustrates the physical picture for the
spatial hypersurfaces.

\section{Conclusion}\label{sec:Conc}
We presented a detailed look into Carter's Kerr-de Sitter
spacetime, which is the uncharged, rotating black hole solution to
Einstein's equation with cosmological constant $\La$. As is known
and was shown here, the $a=M$ case does not yield an extremal
black hole in this spacetime. We quantified this by explicitly
solving the apparent horizon equation for real roots and showing
that the black hole spin tops out around $a \approx 1.1M$. Fig.
\ref{fig:KdS_a_vs_Lambda} demonstrates that for any $\La>0$,
$a_{max} > M$, and if $\La>1/9M^2$ then $a$ cannot be zero.
Furthermore, if $\La$ is larger than some maximum value $\b{\La}
\approx 0.1778/M^2 $ then there is no longer a black hole in the
spacetime. The cosmic expansion tears the black hole apart.  As
far as we know, no one has published computational these results
before.

We also investigated the global structure of the extended solution
by first looking at the correspondence between the $|Q|=M$
Reissner-Nordstr\"{o}m-de Sitter and Kaster-Traschen spacetimes.
Globally, the RNdS spacetime can be viewed as a 3-sphere
containing oppositely charged black/white holes at its poles. In
more detail, we saw that depending on the kind of spacelike
hypersurface, one can get one of the four possible combinations of
$ BH-BH, \ BH-WH, \ WH-BH, \ WH-WH $ at the poles. Using a
coordinate transformation and an approximation scheme, we massaged
the $a=M$ case of KdS to look very much like $|Q|=M$ RNdS and
exactly identical to it in the large $r$ limit. This and the fact
that the $\t = 0 $ cut of KdS and RNdS have the same Penrose
diagrams led us to conclude that KdS has a very similar global
structure to RNdS. In this case, the physical picture (spatially)
is that of a 3-sphere with a Kerr black (or white) hole at each
pole; the Kerr holes counter-rotate. The holes are causally
disconnected: spacelike separated in a nonsingular coordinate
patch. In our description the two hemispheres of the 3-sphere are
covered by a coordinate system made up of two separate natural
coordinate charts and there is a natural stitching boundary at the
null surface $r_C$. A very simple example of this stitching is
given by Brill in \cite{Br} who stitches forward/backward evolving
de Sitter universes (no holes) at the cosmological horizon.

Other matters not yet addressed relating to KdS spacetime include
the frame dragging by the spin of the black holes, and issues with
possible violation of the cosmic censorship conjecture (\cite{Pe})
in this spacetime for the non-generic cases. In \cite{BH} and
\cite{BHKT}, the singularity structure of RNdS spacetime is
investigated in detail. In \cite{BH}, the possible naked
singularities are classified under names like ``generic naked
singularity" or ``extreme naked singularity" depending on which
roots of $f(r) $ in Eq.(\ref{eq:RNdS_f}) are degenerate. The
generic singularity case yields $S^2\times R^1$ topology for the
$t=constant$ spatial hypersurfaces. As before, we can take this to
be a 3-sphere minus the poles. But whereas before the poles were
cloaked behind horizons, now they are simply punctures on the
3-sphere. Since the time coordinates of the hemispheres are time
reversals of each other, the punctures at the poles in RNdS act
like oppositely charged point particles. Although these points are
eternal naked singularities, \cite{NSH} points out that these
singularities do not form from an initial compact spatial
hypersurface (proper initial data) thus obey a form of weak cosmic
censorship. Wald also agrees with this in \cite{Wa2}, but does not
rule out the possibility of RNdS or KdS potentially providing
evidence against strong cosmic censorship.

The case of ``extreme naked singularity" is more interesting
because the RNdS Penrose diagram suggests the creation and
annihilation of a pair of point charges (with total net charge
equaling zero of course, see Fig.1(d) of \cite{BH} and/or Fig.2 of
\cite{BHKT}). These singularities have nonsingular spacelike
surfaces to their past i.e. they develop from regular initial
conditions. Moreover, these singularities are visible from
$\mathcal{I}^+$. However, as shown in \cite{BHKT}, one can find
complete timelike geodesics along which the observer spends an
infinite amount of proper time travelling and never sees the
singularities. According to \cite{BHKT}, these singularities then
are not really `naked' because they are not visible
to all observers originating from a point on a regular spatial 
hypersurface in the past. These difficulties arise from the fact
that there is no clear, universally-agreed-upon definition for
naked singularities when the cosmological constant is non-zero and
there is no asymptotic flatness. These interesting topics are
beyond the scope of this article so we will not say more about
them here.

As our work shows, the Kerr-de Sitter solution has interesting
properties and comes with some puzzling peculiarities. Although at
first, some of these might seem trivial, a closer look shows that
they are anything but. We hope that we brought some clarity to
this picture and helped the reader to have a better understanding
of the Kerr-de Sitter spacetime.

\section{Acknowledgements}
S.A. thanks Leor Barack and David Garfinkle for useful suggestions
and physical insights. He also acknowledges support from STFC
through Grant No. PP/E001025/1.

\appendix
\section{Regularity of the Axis}\label{sec:app1}

In section \ref{sec:KdS}, we noted that the factor of $ \Xi = 1 +
\La a^2/ 3 $ was needed to avoid conical singularities near the
black hole axis. Here, we demonstrate the role this factor plays.
We look at the metric on a section of a spheroid in the vicinity
of its `poles' $\theta=0,\pi $. Near $\theta=0$ , $\sin\theta
\approx \theta << 1$ and $\cos\theta \approx 1 $. Furthermore, for
finite $r$, $ r\sin\theta $ is also small. The relevant part of
the metric is
\be d\Omega_2^2 = g_{\theta\theta}\; d\theta^2 + g_{\phi\phi}
d\phi^2 \label{eq:2metric}. \ee
From Eq.(\ref{eq:KdS_metric1}), near the pole we have
\be g_{\theta\theta} = \f{r^2+a^2\cos^2\theta}{\dt} \approx
\f{r^2+a^2}{\s} \ , \ee
and
\ba g_{\phi\phi} & =& \f{\Delta_\theta \sin^2\theta}{\rho^2
\;\Xi^2}
(r^2+a^2) - \f{\Delta_r a^2 \sin^4\theta}{\rho^2\; \Xi^2}, \nn \\
& \approx & \f{\Delta_\theta \sin^2\theta}{\rho^2\; \Xi^2}
(r^2+a^2), \nn \\ & \longrightarrow & \f{\sin^2\theta}{r^2+a^2}
\f{\s}{(\s)^2} \left(r^2+a^2\right)^2 = \f{r^2+a^2}{\s}
\sin^2\theta . \ea
Above, we made use of the approximation that $\sin^4\theta <<
\sin^2\theta $. The 2-metric of Eq.(\ref{eq:2metric}) becomes
\be d\Omega_2^2 \longrightarrow \f{r^2+a^2}{\s} \left(d\theta^2 +
\sin^2\theta d\phi^2 \right) \label{eq:dS2metric}. \ee
This is the line element on a sphere in the vicinity of its poles.
As can be seen from the 2-metric in Eq.(\ref{eq:dS2metric}), the
factor of $\Xi $ effectively rescales the coordinate radius. For
example, the circumference of a circle near the poles now equals
$2\pi \sqrt{r^2+a^2}(\s)^{-1/2} \sin\theta $ as opposed to just
$2\pi \sqrt{r^2+a^2} \sin\theta $ for a Kerr black hole with
$\La=0$. From a physical perspective, this type of disagreement
with the Kerr result should be expected. For starters, the extra
factor depends on the cosmological constant $\La$; and  $\Lambda$
should have an effect on the size of the black hole. If $\Lambda$
is large, then the black hole would have to be smaller in size in
order to keep itself together against the cosmic repulsion. The
$\Xi^{-1/2}$ dependence of the circumference indicates this
reaction. $\La=0$ gives $\Xi=1$ which corresponds to the regular
Kerr case where we have the usual horizon areal radius
$(r_+^2+a^2)^{1/2}$. As $\Lambda$ grows, $\Xi $ exceeds 1, thus
making the horizon radius smaller than its Kerr value. Basically,
a black hole in a de Sitter universe always has to be smaller than
its Minkowski counterpart in order to hold itself together against
the $\La$ driven expansion that increases for larger $r$.

\section{Surface Gravity of the Kerr-de Sitter Black
Hole}\label{sec:app2} As in the Kerr case, we find a vector field
$\xi^\mu$ that is a geodesic, Killing, tangent to and regular at
the event horizon. If all four conditions are satisfied then the
surface gravity $\kappa$ of the black hole is given by the
following:
\be \xi^\mu \nabla_\mu \xi_\alpha \mid_{EH} = \kappa \xi_\alpha
|_{EH}, \label{eq:geodesicXi} \ee
where $EH$ implies that we must evaluate the equations on the
event horizon. Using Killing's equation $\nabla_\alpha \xi_\beta =
-\nabla_\beta \xi_\alpha$, Eq.(\ref{eq:geodesicXi}) can be
rewritten as
\be -\f{1}{2} \nabla_\mu \xi^2 |_{EH} = \kappa \xi_\mu |_{EH}
\label{eq:geoXi_2} . \ee
The Killing vector tangent to the event horizon of the Kerr-de
Sitter black hole looks very much like its Kerr counterpart (see
\cite{Wa} for a comparison). In Boyer-Lindquist like coordinates
of KdS spacetime, $\xi $ is as follows (\cite{GLPP}):
\be \xi = \Xi \f{\partial}{\partial t} + \Omega_H
\f{\partial}{\partial \phi} \label{eq:KillingXi}, \ee
where
\be \Omega_H = \f{a \; \Xi}{r_+^2+a^2} \label{eq:OmegaH} \ee
is the event horizon angular velocity. Written in Kerr-Schild like
coordinates
the Killing vector becomes
\be \xi = \f{\partial}{\partial \tau} +
\f{a}{r_+^2+a^2}(1-\f{\Lambda}{3} r_+^2) \f{\partial}{\partial
\Phi} \ .  \label{eq:KilXi_KS} \ee
One can readily check that $\xi$ is null on the horizon i.e.
$\xi^2 |_{r=r_+} $. To use the geodesic equation \ref{eq:geoXi_2}
properly, we must use the global form of $\xi^\mu$ where $r_+$ is
left as radial coordinate variable $r$. After all, $\xi^\mu$ is
required to be Killing only on the event horizon. We can compute
the left hand side (LHS) of Eq.(\ref{eq:geoXi_2}) in any
coordinate frame we wish. Using BL coordinates  we get
\be \xi^2  = -\f{\rho^2}{(r^2+a^2)^2} \Delta_r \ . \label{eq:Xi2}
\ee
Here $\Delta_r$ and $\rho^2$ have the usual meanings from section
\ref{sec:KdS}. This expression is a function only of the radial
coordinate $r$, so the only non-zero contribution to the LHS of
Eq.(\ref{eq:geoXi_2}) comes from $\nabla_r =\partial_r$ component:
\be -\f{1}{2} \nabla_\mu \xi^2|_{EH} \longrightarrow -\f{1}{2}
\partial_r \xi^2 |_{r=r_+} = \f{1}{2}\f{r_+^2+a^2\cos^2\theta}{r_+^2+a^2}
\Delta'(r_+) \ . \label{eq:drXi2} \ee
Above, the prime indicates differentiation with respect to $r$ and
we used the fact that $ \xi^2$ is a scalar. After taking the
derivative, we evaluate everything at $r=r_+$. On the right hand
side of Eq.(\ref{eq:geoXi_2}), we have $\xi_\mu$, which has a
non-zero $r$-component because in KS coordinates, the metric has a
$g_{\tau r}$ crossterm. Thus, we have
\be \xi_r = \f{2Mr \rho^2}{(r^2+a^2)^2(1-\lambda r^2)} \ .
\label{eq:Xi_r} \ee
Now, Eq.(\ref{eq:geoXi_2}) becomes
\be \f{1}{2} \f{\rho_+^2}{(r_+^2+a^2)^2} \Delta'_r(r_+) = \kappa
\f{2Mr_+ \rho_+^2}{(r_+^2+a^2)^2(1-\lambda r^2)},
\label{eq:geoXi_3} \ee
which gives for $\kappa$
\be \kappa = \f{1-\f{\La}{3}r_+^2}{4Mr_+} \Delta'_r(r_+) \ ,
\label{eq:kappa} \ee
consistent with Eq.(\ref{eq:expansion3}) above.
Eq.(\ref{eq:kappa}) precisely matches the expression presented in
different notation in \cite{GLPP}.

\section{Kastor-Traschen Spacetime}\label{sec:app3}
First discovered by Kastor and Traschen (\cite{KT}), the KT metric
represents a de Sitter universe containing N black holes, each
located at $\mathbf{R}_i $. For $H\equiv (\Lambda/3)^{1/2} $, the
KT metric is given to be:
\be ds^2 = -\f{d\tau^2}{U^2} + U^2 (dR^2 + R^2 d\Omega_2^2),
\label{eq:KT_metric} \ee
where
\be U = H \tau + \sum_{i=1}^N \f{M}{\left|\mathbf{R}-\mathbf{R}_i
\right|}.
\ee
This is the same metric as that of $|Q|=M$ RNdS spacetime except
with more than one black hole. Physically, this represents the
situation in which the Coulomb force cancels the Newtonian gravity
and the black holes accelerate away from each other under the
influence of the cosmic expansion. As was shown in section
\ref{ssec:RNdS}, the spatial geometry near each black hole is that
of a 3-cylinder. In this ``expanding" coordinate system of
Eq.(\ref{eq:KT_metric}), one starts with N singularities at
$\tau=-\infty$ which then expand into smooth, non-singular spatial
3-cylinders by the time $\tau=0$. For $\tau>0$, we have N black
holes that are smoothly joined near $R=\infty$. The case of a
single mass corresponds to the $|Q|=M$ RNdS metric presented in
section \ref{ssec:RNdS}. One can also write this metric using a
contracting chart.

\end{document}